\documentclass[runningheads]{llncs}
\usepackage{amssymb}
\newcommand{\QEDA}{\hfill \blacksquare}
\usepackage{graphicx}
\usepackage{amsmath}
\usepackage{stmaryrd}
\usepackage{tabularx}
\usepackage{todonotes}
\usepackage{listings}
\usepackage{verbatim}
\usepackage{algorithmic}
\usepackage[ruled,vlined]{algorithm2e}
\renewcommand{\algorithmicrequire}{\textbf{Input:}}
\renewcommand{\algorithmicensure}{\textbf{Output:}}
\usepackage[utf8]{inputenc}
\usepackage{textgreek} 
\usepackage{xcolor}  
\usepackage{placeins}
\usepackage[toc, page]{appendix} 
\usepackage{listings}
\usepackage{float}
\usepackage{comment}
\begin{document}
\title{Verifying Correctness of PLC Software during System Evolution using Model Containment  Approach}
\titlerunning{PLC Software Verification}
\author{
  Soumyadip Bandyopadhyay\inst{1}\and  
  Santonu Sarkar\inst{2}
}
\institute{
  ACM Member\\
  \email{soumyadip.bandyopadhyay@gmail.com}
\and
 BITS Pilani Goa K K Birla Goa Campus, India\\
  \email{santonus@goa.bits-pilani.ac.in}
}
\maketitle
\begin{abstract}
 Upgradation of Programmable Logic Controller (PLC) software is quite common to accommodate evolving industrial requirements. Verifying the correctness of such upgrades remains a significant challenge. In this paper, we propose a verification-based approach to ensure the correctness of the existing functionality in the upgraded version of a PLC software. The method converts the older and the newer versions of the sequential function chart (SFC) into two Petri net models. We then verify whether one model is contained within another, based on a novel containment checking algorithm grounded in symbolic path equivalence. For this purpose, we have developed a home-grown Petri net-based containment checker.
Experimental evaluation on 80 real-world benchmarks from the OSCAT library highlights the scalability and effectiveness of the framework. We have compared our approach with \texttt{verifAPS}, a popular tool used for software upgradation, and observed nearly 4x performance improvement. 
\end{abstract}
%
%
%
\keywords{Petri net, PLC software verification,  industrial automation, Sequential Function Chart (SFC),  containment checking, software evolution.}
\section{Introduction}
A modern automation system is driven by process control software written in various programmable logic controller (PLC) languages. To remain efficient, reliable, and sustainable, existing software must be continually upgraded to meet evolving market demands and new safety regulations. However, modifying PLC software carries far more risks, particularly unintended regressions may disrupt existing process operations, leading to disastrous consequences. Therefore, ensuring the correctness of upgraded PLC software is extremely critical. 

Traditional approaches to PLC software regression testing primarily rely on simulation and regression tests, which can be time-consuming, incomplete, and dependent on domain-specific expertise. While regression testing is commonly employed to compare outputs before and after software modifications, it may not always capture behavioral discrepancies introduced during upgrades. 

In this paper, we propose a novel verification-based approach that formally ensures the correctness of existing functionality in the upgraded PLC software. Here, we consider a popular IEC 61131-3 compliant PLC language known as the Sequence Function Chart (SFC). We transform both the older and newer versions of the SFC into Petri net models and verify whether one model is contained in the other. This containment check is based on the notion of behavioral equivalence, ensuring that the upgraded version preserves the expected execution behavior of the previous version. The contributions of our paper are as follows.
\begin{enumerate}
\item Model Enhancement- We present an improved version of the Petri net model, where a place can have a sequence of functions (instead of a single function in\cite{souban,ppl}), while transitions are now limited to guard conditions only. This refinement leads to a more compact model representation. We further propose the notion of {\em execution cut point} to handle PLC tick semantics.
   \item Petri Net-based Model Construction – We introduce a model transformation process that converts an SFC program into a Petri net in order to compare behaviors across versions. The transformation process handles synchronous execution semantics through tick labeling, which is crucial for PLC verification but not addressed in general Petri net approaches.
   \item Containment Checking Tool – We defined a new theoretical foundation of {\em containment} and proposed an algorithm to check containment of one Petri net in another. This, in turn, verifies behavioral consistency between different software versions. 
    \item Scalability and Practical Evaluation – We validate our approach on a diverse set of eighty benchmarks from the open-source OSCAT library, demonstrating its effectiveness in real-world PLC software evolution scenarios. We have also compared our approach with the popular open-source {\tt verifAPS} tool.
\end{enumerate}
The paper has been organized as follows. Section~\ref{sec:ex} introduces a motivating use-case as a running example in the paper. Section~\ref{sec:funcarch} describes the functional architecture of the proposed tool. Section~\ref{sec:sfc2pn} illustrates the construction of a Petri net model. We describe the containment checking process in Section~\ref{sec:contchk}. We report the experimental evaluation of the tool in Section~\ref{sec:expResult}. We discuss the utility and limitations of our approach in Section~\ref{sec:discussion}. Section~\ref{sec:rw} discusses several important relevant contributions in this area. Finally, we conclude our paper.

\section{Motivating Example}\label{sec:ex}
\begin{figure}
    \centering
   \includegraphics[width=\textwidth]{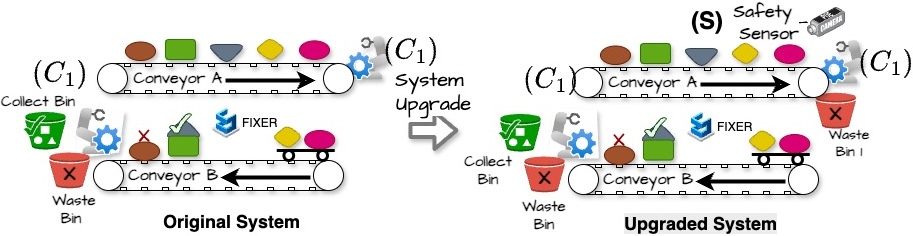}
     \caption{Pick and Place System- original and upgraded version with safety features}
    \label{fig:funcarch}
    \vspace{-0.2in}
 \end{figure}
Within this paper, we employ a widely recognized ``Pick and Place system''~\cite{unknown} as an ongoing example. The system outlined is a simplified representation of a complex industrial process where items are initially placed on Conveyor A. A robotic arm then simultaneously transfers two items from Conveyor A to Conveyor B. Conveyor B is equipped with a scaling sensor that verifies the scaling parameters of these items, denoted as $a$ and $b$. When these parameters are positive, a fixer combines them, and a second robotic arm deposits the composite item into the green \textit{Collect Bin}. Conversely, if the parameters fail to meet a positive value, the items are put to the red \textit{Waste Bin}. The fixing process hinges on intricate calculations adjusting individual scaling factors of items $a$ and $b$ to derive an integrated, custom scaling metric. 

In its upgraded version, the system includes a newly installed safety guard sensor, complying with the latest safety protocol (ISO 10218-1)\footnote{https://www.iso.org/standard/73933.html}. This sensor oversees the overall system operation and ensures that, without the safety guard being active, the robotic arm on Conveyor A remains immobilized, preventing any object transfer from Conveyor A to B due to breach of safety norms.
If the safety guard is {\em in place}, the robot will place the objects on the conveyor belt B, and the scaling sensor will check the scaling parameters as in the original version.
Furthermore, the upgraded version computes the fine-tuning of the individual scaling factors (for $a$ and $b$) concurrently.
The classical testing method is not useful for two reasons.
\begin{enumerate}
    \item An SFC software testing generally takes 3 to 5 minutes by an expert engineer in a testing environment.
    \item The older version is purely sequential, whereas the upgraded version has parallelism, as it performs the scaling computation concurrently.
\end{enumerate}

The entire control logic is implemented in SFC. The SFC for Figure \ref{fig:SFC}(a) is in Figure \ref{fig:PN}(c), and for the upgraded version in Figure \ref{fig:SFC}(b), it is shown in Figure \ref{fig:PN}(d).

\section{Containment Checking Software}\label{sec:funcarch}
\vspace{-0.2in}
\begin{figure}
    \centering
   \includegraphics[width=\textwidth]{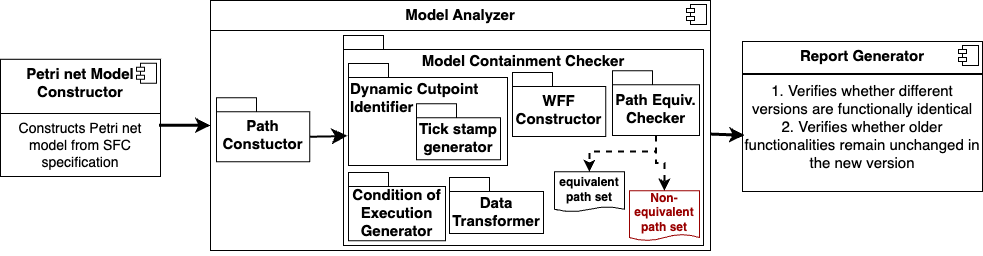}
     \caption{Functional Block Diagram}
    \label{fig:funcarch1}
    \vspace{-0.2in}
\end{figure}
The containment verification tool illustrated in Figure~\ref{fig:funcarch1} includes three key components: the Petri net model constructor, the Model Analyzer, and the Report Generator. As detailed in Section~\ref{sec:sfc2pn}, the Petri net model constructor transforms an SFC program into a Petri net model. {\em Petri nets are selected as the formal model due to their semantics being well-aligned with those of Sequential Function Charts (SFC) \cite{406973}, facilitating a more intuitive and precise translation. Conversely, converting SFC to a Control Data Flow Graph (CDFG) entails multiple intermediate transformations \cite{alexander1}, increasing the modeling complexity.} The Path constructor component generates an array of paths derived from a set of cut-points to a cut-point without any intermediary cut-points.
The Model Containment Checker component takes two Petri net models—one for the original SFC code and the other for the updated SFC code—to determine if the updated SFC retains the original SFC's functionality. This process is referred to as model containment checking (Definitions~\ref{d:eqModels} and ~\ref{d:contModels}). This component leverages data transformation and an execution condition checker to assess model containment. We incorporate the concept of path equivalence checking~\cite{souban} for this purpose. The report generator module delivers a comprehensive textual explanation of how two SFC codes meet the containment requirements, and if there is a violation, it explains why the containment relationship does not hold. 
\section{Petri net based Containment Checking}\label{sec:sfc2pn}
\vspace{-0.3in}
 \begin{figure}
    \centering
     \includegraphics[width=\textwidth]{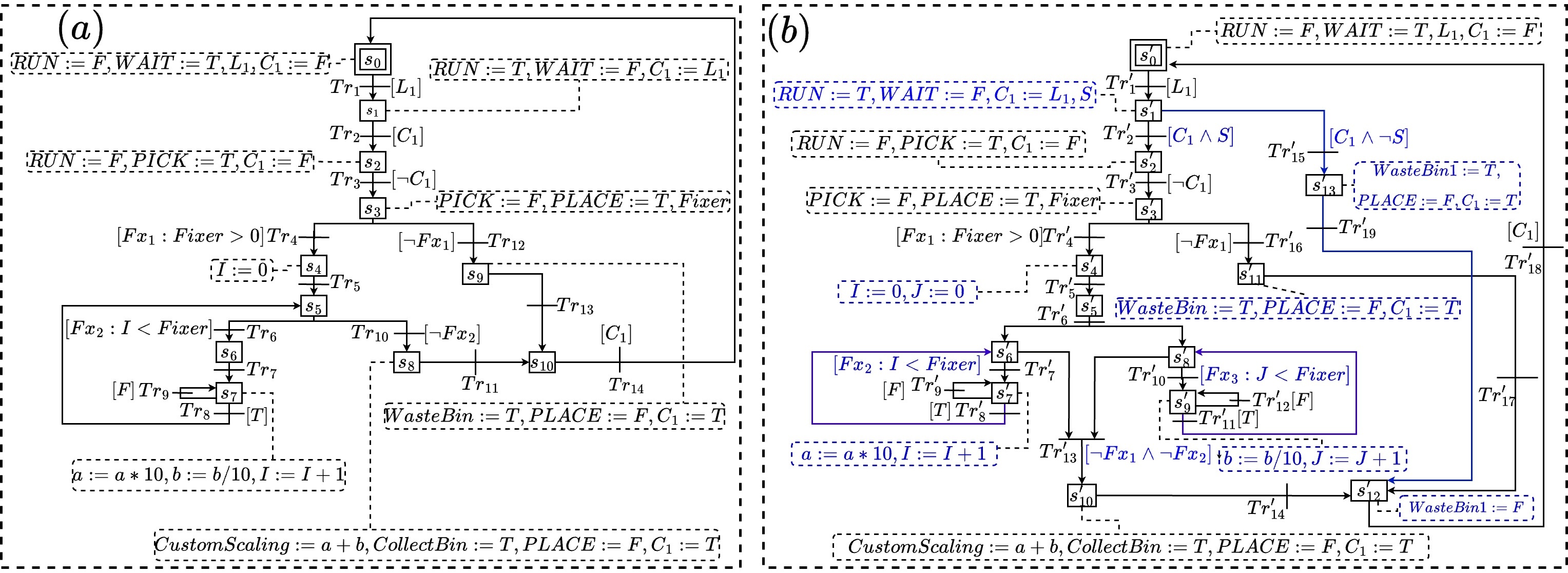}
     \caption{Two versions of Pick and Place, (a) Original SFC (b) Upgraded SFC (changes shown in blue color)}
    \label{fig:SFC} 
    \vspace{-0.2in}
 \end{figure}
An SFC code is comparable to a Petri net; formally described in \cite{BauerHLE04}. We reproduce the definition for brevity.
\begin{definition}
A sequential function chart (SFC) $\mathcal{S}=\langle S, X, A, T_S, s_0 \rangle$, is a structure where $S$ is a finite set of steps, $X$ is a finite set of variables, $A$ is a finite set of actions,
$T_S$ is a finite set of transitions for SFC, and $s_0$ is the initial step.
\end{definition}
Let $\Sigma_{SFC}$ be the all possible states for variable in $X$, i.e.,
$ \Sigma_{SFC} = \{\sigma| \sigma:  X \rightarrow Value\}$ where $\sigma$ assigns a $val\in Value$
to each variable in $X$. Next, $a_N \in A$ is an action, modeled as a state transformation function $\mathcal{T}: \Sigma_{SFC}  \rightarrow \Sigma_{SFC}$.  

A \texttt{block} is an action labeling function that assigns to each step a collection of pairs $\langle a_N, q \rangle$, referred to as an action block, where $a_N$ is an action and $q \in$ $\{entry, active, exit\}$ denotes an action qualifier.
 We consider the qualifiers \emph{entry}: to indicate computation during the entry of the step, \emph{exit:} to indicate computation during the exit of the step, and \emph{active:} to indicate computation when the step is active. 
There is a flow relation between a \emph{step} to a \emph{transition} and a \emph{transition} to a \emph{step}. Each transition $t \in T_S$ is associated with a {\em guard condition} $g_s$ which is a boolean function over a subset of the variables in $X$.

SFC semantics contains a global time, modeled by a {\em tick}\footnote{https://control.com/technical-articles/an-overview-of-iec-61131-3-Industrial-Automation-Systems/}~\cite{LeeB24}. Computation in the {\em active} step happens at each {\em tick} value. After the computation, the {\em tick} value is incremented. 
We do not consider hierarchical SFCs in this work, where actions may also contain another SFC.
\begin{example}\label{eg:SFC}
Figures~\ref{fig:SFC}(a) and ~\ref{fig:SFC}(b) are SFC programs for the original and the upgraded version of the pick and place system (Figure~\ref{fig:funcarch}), respectively. In Figure \ref{fig:SFC}(a), $S=\{s_{0},\cdots,s_{10}\}$, $T_S=\{Tr_{1},\cdots,Tr_{14}\}$. Every step $s\in S$ is associated with an action sequence. The action sequence for $s_{7}$ is {\tt a= a*10}, followed by {\tt b=b/10} and then {\tt I++}.
Every transition has a guard condition. In Figure \ref{fig:SFC}(a), the guard condition associated with $Tr_{4}$ is $F_{x_1}$  and $Tr_{12}$ is $\neg F_{x_1}$. Here $F_{x_1}$ denotes the expression $Fixer>0$. If no guard condition is explicitly associated with a transition, it is treated as {\em true}. 

In Figure \ref{fig:SFC}(a), the SFC starts from step $s_0$, with the action \texttt{WAIT=T}, while all other actions are false. The sensor $L_1$ represents a load sensor on conveyor belt A. When an object is detected on this conveyor, $L_1$ becomes true, enabling transition $Tr_1$ and progressing to $s_1$, with update actions \texttt{RUN=T, WAIT=F, and $C_1=T$}.

Next, with the activation of the guard condition $C_1=T$, transition $Tr_2$ is enabled, leading to $s_2$. Here, the robotic arm picks the object from conveyor belt A, and $C_1$ is set to false (\texttt{RUN=F, PICK=T, and $C_1=F$}).

When the guard condition for transition $Tr_3$ is true, the system enters $s_3$ where the object is transferred from conveyor belt A to B, and the action \texttt{PLACE=T} is set. During this placement, the robotic arm also checks the scaling parameter value using a sensor attached to the arm. 
If the scaling parameter is positive, they are fixed by a fixer, via step $s_4$ and the object is placed in the collection bin at $s_8$. Otherwise, the object is sent to the waste bin through step $s_9$.

In the upgraded version shown in Figure \ref{fig:SFC}(b), an additional safety sensor ($S$) monitors the system's status.
If the safety guard is {\em false}, the robot arm of Conveyor A remains locked, preventing the robot from transferring any object from Conveyor A to B.
Instead, all objects are redirected to the waste bin via step $s'_{13}$. Additionally, the custom scaling process is now carried out in parallel. The blue-colored portion in Figure \ref{fig:SFC}(b) indicate the upgraded portions of the SFC.
$\QEDA$

\end{example}

\begin{definition}
A Petri net model is a $7-$ tuple $N = \langle P, V, F, T, I, O, P_{M_0} \rangle$, where 
 $P$ is a finite non-empty set of places, $V$ is a set of variables, 
 $F$ be set of all update functions, $T$ is a finite non-empty set of transitions,
 $I \subset P \times T$ is a finite non-empty set of input arcs which define the flow relation between places and transitions, $O \subset T \times P$ is a finite non-empty set of output arcs which define the flow relation between transitions and places and $P_{M_0}$ is the initial set of place marking.
\end{definition}
Each place is associated with a set of variables using the function $\phi:P\to2^V$. Each element of $V$ belongs to one of two disjoint subsets, namely the set of changed variables
($V_c$) or the set of unchanged variables ($V_u$). Here we only consider integer and boolean types.
Each place is associated with a sequence of update functions $F_p \subseteq F$ for the variables in $V$ where each function is of the form $\Sigma_{PN} \rightarrow \Sigma_{PN}$ such that $\Sigma_{PN}$ represents set of all possible states of variable in $V$. 

A place can hold a \textit{token}. A place with a token is called a {\em marked place}. 
 When a place is \textit{marked}, it computes a sequence of functions associated with it. The corresponding values of the variables, as defined by these functions, are updated. 

Each transition $t$ is associated with a {\em guard condition} $g_t$, a boolean function over a subset of the variables. 
A transition $t$ is said to be {\em enabled} when all its pre-place(s) have token(s), and they are associated with the set(s) of values of the variable which satisfy $g_t$. All the enabled transitions can fire simultaneously, which in turn models parallel execution. 
 A transition $t_s \in T$ is said to be a \emph{synchronizing transition} if its post-places are all the initial marked places. 
Our Petri net model is deterministic and 1-safe (at any point in time, a place contains at most one token, but one token corresponds to k variables).
\begin{figure}
    \centering
     \includegraphics[width=\textwidth]{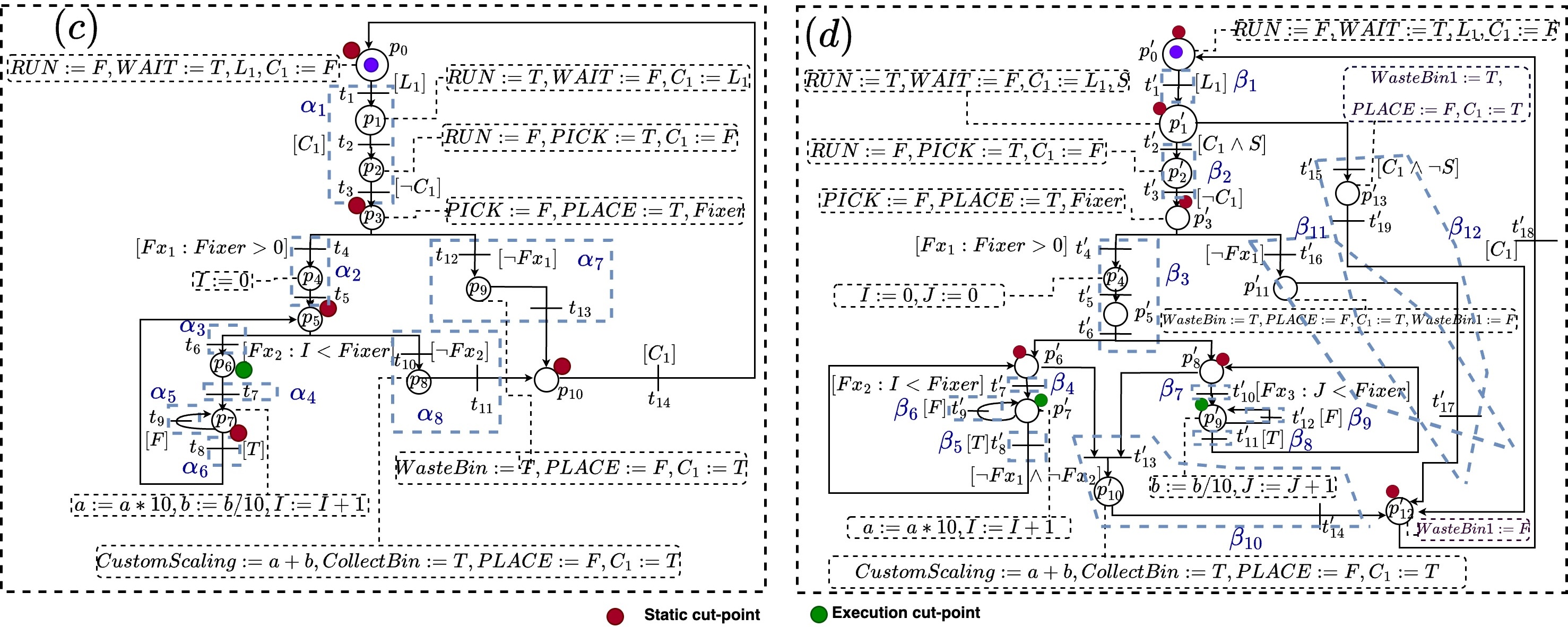}
     \caption{Two versions of Petri nets}
    \label{fig:PN}
    \vspace{-0.4in}
\end{figure}
\subsection{SFC to Petri net translation}
Translating an SFC into a Petri net is syntactic and straightforward, where one can generate a Petri net in a single pass over the SFC structure. The Petri net model's entities $I$ and $O$ correspond to the SFC's flow relation. The mapping of SFC entities to Petri net entities is $S\to P$, $X\to V$, $T_s\to T$, $g_s\to g_t$, $s_0\to P_{M_0}$, and $A\to F$. Note that the action names for each action block in the SFC are executed in the same order as the functions in $F_p$ in the Petri net, with both following the ordering defined by the action qualifier in the order of {\em entry}, {\em active}, and {\em exit} for the particular action block. 

For any \textit{state-modifying action} (e.g., incrementing a variable) associated with a step in a Sequential Function Chart (SFC), the corresponding Petri net sub-net can be modeled as follows:

\begin{enumerate}
    \item A \textit{place} \( p \) represents the corresponding \textit{SFC step} in the Petri net.
    \item This place is connected to two \textit{mutually exclusive transitions}.
    \item One of these transitions, denoted \( t_u \), models the execution of the state-modifying action. It forms a self-loop with \( p \), represented by the flow relations \( (p, t_u) \) and \( (t_u, p) \).
    \item The transition \( t_u \) is enabled under the \textit{negation of the condition} that enables the alternative transition.
    \item All enabled transitions in the Petri net contain the same {\em tick} value
\end{enumerate}


\subsection{Computation and Containment of Petri net model}
Before describing the containment checking in this subsection, we describe the prerequisites of the Petri net model related to computational equivalence. 
%
\begin{definition}[Successor marking]\label{d:succmarking}
A set of marked places $P_{M^{+}}$ is said to be a successor place
marking of $P_{M}$, if 
$P_{M^{+}}$ 
includes all post-places of transitions \( T_M \) enabled by \( P_M \). and also all the places of $P_M$ whose post-transitions are not enabled; symbolically, 
$P_{M^+} = \{p \mid p \in$ $t^{\circ}$ and 
$t \in T_{M}\} \cup \{p \mid p \in P_{M}$ and $p \notin$ $^{\circ}T_{M}\}$.

\end{definition}
\begin{definition}[Computation]\label{d:comp}
 In a Petri net model $N$ with initial marked places $P_{M_0}$, 
a computation $\mu_{p}$ of an out-port $p$ (place having a synchronizing transition) is a sequence  
$\langle T_1, T_2, \ldots, T_i, \ldots, T_l \rangle$ of 
sets of maximally parallelizable transitions \cite{souban} satisfying the following 
properties:
\begin{enumerate}
  \item There exists a sequence of markings of places 
         $\langle P_{M_0}, P_{M_1}, \ldots, P_{M_{l-1}}\rangle$ such that
   \begin{enumerate}
    \item $P_{M_0}$ be the set of initial marked places
    \item $\forall i, 1 \leq i < l$, $P_{M_i}$ is a successor place marking of 
         $P_{M_{i-1}}$, $^{\circ}T_{i} \subseteq P_{M_{i-1}}$ 
          and $T_{i}^{\circ}\subseteq P_{M_i}$.
   \end{enumerate}
  \item $p \in T_{l}^{\circ}$.
\end{enumerate}
\end{definition}
There are two entities associated with every computation: 
(1) the condition of execution denoted as $R_{\mu_p}$ and (2) the data transformation denoted as $r_{\mu_p}$; both are expressed in normalized form \cite{ds}.
\begin{definition}[Equivalence of a Computation]\label{d:eqComputations}
Let $N_0$ and $N_1$ be two Petri net models with their set of initial place marking bijection $f_{in}$ and out-port
     bijection $f_{out}$.
A computation $\mu_p$ of a model $N_0$ having an out-port $p$ is said to be equivalent 
to a computation $\mu_{p'}$ of another model $N_1$ having an 
out-port $p'$ such that $f_{out}(p) = p'$, symbolically denoted as $\mu_p \simeq_c \mu_{p'}$, 
if $R_{\mu_p} \equiv R_{\mu_{p'}}$ and  $r_{\mu_p} = r_{\mu_{p'}}$. 
\end{definition}
%
%
\begin{definition}[Containment of two models]\label{d:contModels}
The Petri net model $N_0$ is said to be contained in the Petri net model
$N_1$, represented as $N_0 \sqsubseteq N_1$, 
if, $\forall p \in$ out-port of $N_0$, for any computation $\mu_{p}$ of $p, \exists$ a computation $\mu_{p'}$ of an out-port $p'=f_{out}(p)$ of $N_1$ 
such that $\mu_{p} \simeq_c \mu_{p'}$.
\end{definition}
\begin{definition}[Equivalence of two models]\label{d:eqModels}
A model $N_0$ is said to be computationally equivalent to a model $N_1$, symbolically denoted as
$N_0 \simeq N_1$, if $N_0 \sqsubseteq N_1$ and $N_1 \sqsubseteq N_0$.
\end{definition}

In symbolic computation-based program equivalence checking, a loop is cut to create a cut-point. Then, a path $\alpha$ is constructed from one cut-point to another without any intermediary cut-points such that \textit{any computation can be represented as a concatenation of paths}~\cite{Manna71,ne,amir,mit}. 

The notion of cut-points and paths has been adopted in Petri-net based models~\cite{souban,ppl} as \textit{static cut-points} for program equivalence checking. We revisit these concepts here since our approach is also built upon the general notion of cut-points and paths.
Static cut-points are introduced in initial marked places, places with multiple post-transitions, places containing back edges, and out-ports. This enables us to break the model into a finite set of paths between these cut-points. The symbol $^{\circ}\alpha$ denotes the set of places where the path $\alpha$ originates. 

Like computation (Definition~\ref{d:comp}), every path $\alpha$ is associated with two entities: 1) the condition of execution along the path, denoted as $R_{\alpha}$ and 2) the data transformation along the path $r_{\alpha}$. For example, consider Figure~\ref{fig:PN}(c). For the path $\alpha_1$, $R_{\alpha_1}=L_1\rightsquigarrow C_1\rightsquigarrow \neg C_1$, and $r_{\alpha_1}=$
\texttt{\{RUN=F, PICK=F, PLACE=T, WAIT=F, $C_1=F$,Fixer\}}.
 The WFF constructor component (Figure~\ref{fig:funcarch}) stores these expressions in the normalized form \cite{ds}.


While a static cut-point is a handy tool to model a computation, it is not sufficient when a computation involves an unknown number of loop traversals, as explained Example~\ref{eg:pathdcp}.
\begin{figure}
\centering
\includegraphics[width=\textwidth]{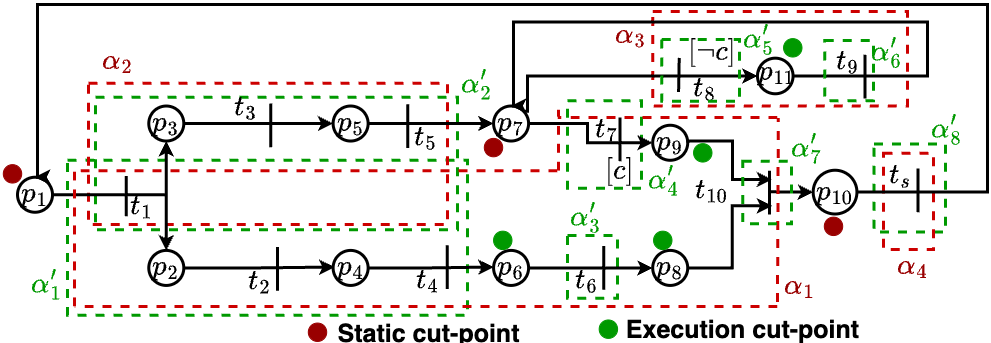}
 
\caption{A Petri net model of a computation involving loops}
\label{fig:counter}
\vspace{-0.2in}
\end{figure}
\begin{example}\label{eg:pathdcp}
Let us consider the computation in Figure \ref{fig:counter}. The set of static cut-points is
$\{p_1, p_7,p_{10}\}$
and the paths will be $\alpha_1= 
\langle \{t_1\},\{t_2\},\{t_4\},\{t_6\}, \{t_7\} \rangle$,  
 $\alpha_2 = \langle \{t_1\},\{t_3\}, \{t_5\}\rangle$,  $\alpha_3 = \langle \{t_8\},\{t_9\}\rangle$ and $\alpha_4 =\langle \{t_s\}$.
Let us consider a computation $\mu$ of the out-port $p_{10}$, where 
$\mu = 
\langle T_1 = \{t_1\}, T_2 = \{t_2, t_3\}, T_3 = \{t_4, t_5\}, 
T_{4}=\{t_6, t_8\}, (T_{5}=\{t_9\}, T_6=\{t_8\}, T_5 =\{t_9\})^{n}, T_7=\{t_7\}, T_8=\{t_{10}\} \rangle$. Since $t_{10}$ (in $\alpha_1$) and $t_8$ (in $\alpha_3$) are mutually exclusive transitions and can't execute in parallel, $\alpha_1$ and $\alpha_3$ need to be concatenated in any computation, hence in  $\mu$ as well. However,
\emph{There does not exist any concatenation of paths $\{\alpha_1, \alpha_2, \alpha_3\}$ that can express this computation $\mu$ due to the transition group $T_4$}. $T_4$ has transitions $\{t_6, t_8\}$ which execute in parallel. However, $t_6$ is a part of the path $\alpha_1$, and $t_8$ belongs to $\alpha_3$. As a result, $\alpha_1, \alpha_3$ cannot be concatenated in any order.

However, if we had $p_6, p_7, p_8,p_9$ and $p_{11}$ also as cut-points,
the path-set would have been $\alpha_{1}^{'} = \langle \{t_1\} , \{t_2\},\{t_4\}\rangle, \alpha_{2}^{'} = \langle \{t_1\}.\{t_3\}.\{t_5\} \rangle,
\alpha_{3}^{'} = \langle \{t_6\}\rangle, \alpha_{4}^{'} = \langle \{t_7\} \rangle, 
\alpha_{5}^{'} = \langle \{t_8\} \rangle,
\alpha_{6}^{'} = \langle \{t_9\} \rangle$ and $\alpha_{7}^{'} = \langle \{t_{10}\} \rangle$\
(green boundaries in Figure \ref{fig:counter}). 
Now, intuitively, the computation $\mu$ can be depicted as the sequence 
$(\alpha_{1}^{'} \parallel \alpha_{2}^{'}).(\alpha_{3}^{'} \parallel \alpha_{5}^{'}).
(\alpha_{6}^{'}. \alpha_{5}^{'})^{n}.
\alpha_{4}^{'}.\alpha_{7}^{'}$ 
of concatenation of parallelizable paths from the set
$\{\alpha_{1}^{'}, \alpha_{2}^{'}, \alpha_{3}^{'}, 
\alpha_{4}^{'},$ 
$\alpha_{5}^{'}, \alpha_{6}^{'}, \alpha_{7}^{'}\}$, where
$(\alpha_1 \parallel \alpha_2)$ means parallel execution of $\alpha_1$ and $\alpha_2$ and 
$(\alpha_1 . \alpha_2)$ means sequential execution
$\alpha_1$ followed by $\alpha_2$. $\QEDA$
\end{example}
The concept of \textit{Dynamic cut-points} was introduced~\cite{ppl} to overcome the drawbacks associated with static cut-point based computation, as demonstrated in Example~\ref{eg:pathdcp}. Although dynamic cut-points are more suitable for our use case, they were originally crafted for procedural languages such as C. In the context of synchronous reactive systems like SFC, which feature the concept of \textit{ticks}, we have expanded the dynamic cut-point framework to include ticks. We call this the \textit{Execution cut-point}, defined below.
\begin{definition}[Execution cut-point]\label{d:dy}
A place $p$ is designated as a Execution cut-point if, during a token tracking execution
of the model~\cite{Wisniewski2020} (with static cut-points already incorporated), a place marking $P_M$ containing $p$ 
is encountered such that one of the following conditions is satisfied:
\begin{enumerate}
 \item $P_M$ contains at least one cut-point; or 
 \item $P_M$ contains more places than its pre-transitions $\mathcal{T}_M$ whose tick values are identical, i.e., 
 $|P_M| > |^{\circ}P_M| \wedge $
 ${}^{\circ}P_M = \left\{ t \in \mathcal{T}_M \mid \text{tick}(t) = v \right\}, \quad \text{where } v$ \text{ is a fixed tick value common to all } $t \in \mathcal{T}_M$; this indicates that the token tracking execution has reached a point of creation of some parallel threads. 
 \end{enumerate}
\end{definition}
%
Henceforth, both static and execution cut-points will be collectively referred to as cut-points for simplicity. 
\begin{definition}[Path cover]
A finite set of paths
$\Pi$ $= \{\alpha_0, \alpha_1, \ldots, \alpha_k\}$ is said to be a path cover of a Petri net model $N$
if any computation $\mu$ of an out-port of $N$ can be represented as a sequence of concatenations of 
parallelizable paths from $\Pi$.
\end{definition}
The definition of the parallelizable path and the concatenation of a path, along with its characteristic, is reported in \cite{ppl}.
\begin{definition}[Place, Transition and Variable correspondence]\label{d:corP}
  Let $N_0$ and $N_1$ be two Petri net models with their initial marked places bijection $f_{in}$ and out-port
     bijection $f_{out}$. 
     Equivalence of paths of $N_0$ and $N_1$, 
     a transition correspondence relation,
     denoted as $\eta_t \subseteq T_0 \times T_1$, and a place correspondence relation, denoted as 
     $\eta_p \subseteq P_0 \times P_1$, are defined as follows:
       \begin{enumerate}
           \item  $f_{in} \subseteq \eta_p$,
           
           \item Two paths $\alpha$ of $N_0$ and $\beta$ of $N_1$ are said to be equivalent, denoted as 
           $\alpha \simeq \beta$, if
           $\forall p \in {}^{\circ}\alpha$, there exists exactly one $p' \in {}^{\circ}\beta$ such that
           $\phi^{0}(p) = \phi^{1}(p')$, 
           $\langle p,p' \rangle \in \eta_p$, 
           $R_{\alpha}(\phi^{0}(^{\circ}\alpha)) \equiv R_{\beta}(\phi^{1}(^{\circ}\beta))$,
           $r_{\alpha}(\phi^{0}(^{\circ}\alpha)) = r_{\beta}(\phi^{1}(^{\circ}\beta))$ and 
           \textit{TickStamp(last($\alpha$))}=\\
           \textit{TickStamp(last($\beta$))}. Note that \texttt{last($\alpha$)} indicates the last member of the path $\alpha$.
           
           \item For any two equivalent paths $\alpha, \beta$, $\langle$ last($\alpha$),
           last$(\beta)$ $\rangle \in \eta_t$ if their \textit{TickStamps} are identical.
           

\item
Let $V_0$ and $V_1$ be the set of variables for $N_0$ and $N_1$ respectively, and the variable correspondence relation $\eta_v$ is defined as $\eta_v \subseteq ((V_0-V_1) \times V_1)\cup(V_0 \times(V_1-V_0)$; thus, if $\langle v, v'\rangle\in \eta_v$ then 
either $(v\in V_0)\wedge(v\notin V_1)$ or, $(v'\in V_1)\wedge(v'\notin V_0)$. Furthermore,
for any two equivalent paths $\alpha$ of $N_0$ and $\beta$ of $N_1$, for any two places $p \in$ post place of $\alpha, p' \in$ post place of path $\beta, 
\langle p, p' \rangle \in \eta_p$ if
 \begin{enumerate}
   \item $\phi^0(p) = \phi^1(p')$ or
   \item if $\phi^0(p) \in V_0 - V_1$ or 
                  $\phi^1(p') \in V_1 - V_0$ then  
     \begin{enumerate}
        \item  either  
                     $\langle \phi^0(p),    
                      \phi^1(p') \rangle$ have  
                     already been associated with    
                     each other in $\eta_v$,
                     or 
      \item  they have  not yet been  
                   associated with any other 
                   variables and can now be 
                   associated associated in
                   $\eta_v$; also $\langle p, p'\rangle$ is put in $\eta_p$;
    \end{enumerate}
 \end{enumerate}
\end{enumerate}  
\end{definition}
\begin{theorem}
A Petri net model $N_0$ is contained in another
Petri net model $N_1$, denoted as $N_0 \sqsubseteq N_1$, for all finite path cover $\varPi_0 =
\{\alpha_{0}, \alpha_{1}, \ldots, \alpha_{l}\}$ of $N_0$ for which there exists a set $\varPi_1 =
\{\beta_{0}, \beta_{1}, \ldots, \beta_{l}\}$ of paths of $N_1$ such that for all $i, 0 \leq i \leq l,$
$(i)$ $\alpha_{i} \simeq \beta_{i}$, $(ii)$ the places in $^{\circ}\alpha_{i}$ have correspondence with those in $^{\circ}\beta_{i}$
and $(iii)$ the places in $\alpha_{i}^{\circ}$ have correspondence with those in 
 $\beta_{i}^{\circ}$.
\end{theorem}
The detailed proof is given in the Appendix.

\section{Containment checking method}\label{sec:contchk}
We  describe the containment checking approach in Algorithm~\ref{Algo:veriDCP}.
\begin{algorithm}
\footnotesize
\caption{ContainmentChecker($N_0, N_1$)} 
\label{Algo:veriDCP}
\begin{algorithmic}[1]
\REQUIRE Petri net models $N_0$ and $N_1$
\ENSURE A six-tuple: 
(1) $\varPi_0$: path cover of $N_0$, 
(2) $\varPi_1$: path cover of $N_1$ matching $\varPi_0$, 
(3) $E$: set of $\langle \alpha, \beta \rangle$ such that $\alpha \simeq \beta$, 
(4) $\eta_t$: matched transition pairs, 
(5) $\varPi_{n,0}$: unmatched paths from $N_0$, 
(6) $\varPi_{n,1}$: unmatched paths from $N_1$

\STATE $\eta_p \gets \{\langle p, p' \rangle \mid p \in P_{M_0} \land p' \in P'_{M_0} \land p' = f_{in}(p)\}$;~~ 
 $\eta_t \gets \emptyset$
\STATE $\varPi_0'\gets \textbf{PathConstructor}(N_0)$; ~~$\varPi_1' \gets \textbf{PathConstructor}(N_1)$
\STATE Initialize $\varPi_0, \varPi_1, \varPi_{n,0}, \varPi_{n,1}, E$ to $\emptyset$

\FORALL{$\alpha \in \varPi_0'$}
\STATE Compute $R_{\alpha}$ and $r_{\alpha}$

    \STATE $\Gamma' = \textbf{SelectedPathForCheckingEquivalence}(\alpha, \varPi_1', \eta_t, f_{in})$
    \FORALL{$\beta \in \Gamma'$}
    \STATE Remove all uncommon variables and then Compute $R_{\beta}$ and $r_{\beta}$
        \IF{$(R_{\alpha} \simeq R_{\beta}) \land (r_{\alpha} == r_{\beta}) \land (\text{TickStamp}(\text{last}(\alpha)) == \text{TickStamp}(\text{last}(\beta)))$}
            \STATE $\eta_t = \eta_t \cup \{\langle \text{last}(\alpha), \text{last}(\beta) \rangle\}$
            \STATE $E = E \cup \{\langle \alpha, \beta \rangle\}$
            \STATE $\varPi_0 = \varPi_0 \cup \{\alpha\}$; $\varPi_0' = \varPi_0' \setminus \{\alpha\}$
            \STATE $\varPi_1 = \varPi_1 \cup \{\beta\}$; $\varPi_1' = \varPi_1' \setminus \{\beta\}$
            \STATE $\eta_p = \eta_p \cup \{\alpha^{\circ}, \beta^{\circ}\}$ \COMMENT{$\beta \simeq \alpha$}
        \ELSIF{(($R_{\alpha_1}\lesssim R_{\beta_1}) \land$ (\text{TickStamp}(\text{last}$(\alpha)) < $\text{TickStamp}(\text{last}($\beta$))) $\vee(r_{\alpha} = \emptyset$))  }
            \STATE $\varPi_0 = \textbf{prepareForExtension}(\alpha, \varPi_0,\varPi_{n,0}, \eta_t, E)$ // extend $\alpha$
        \ELSIF{(($R_{\beta_1}\lesssim R_{\alpha_1}) \land (\text{TickStamp}(\text{last}(\alpha)) > \text{TickStamp}(\text{last}(\beta)))$ $\vee (r_{\beta} = \emptyset))$}
            \STATE $\varPi_1 = \textbf{prepareForExtension}(\beta, \varPi_1, \varPi_{n,1}, \eta_t, E)$ // extend $\beta$
        \ELSIF{$(R_{\beta} \simeq R_{\alpha}) \land (\text{TickStamp}(\text{last}(\alpha)) \neq \text{TickStamp}(\text{last}(\beta))) \land (r_{\alpha} \neg = r_{\beta})$}
            \STATE $\varPi_0' = \textbf{prepareForMerging}(\Gamma)$ \\//merging operation through $N_0$ or $N_1$
        \ELSIF{$(R_{\alpha} \neg \simeq R_{\beta})$}
            \STATE $\varPi_{n,0} = \varPi_{n,0} \cup \{\alpha\}$ 
            \STATE $\varPi_{n,1}=\varPi_{n,1} \cup \{\beta\}$
            \STATE $\varPi'_0 = \varPi_0 \setminus \{\alpha\}$ //Non-equivalent
            \STATE $\varPi'_1= \varPi_1 \setminus \{\beta\}$ //Non-equivalent
        \ENDIF
        
    \ENDFOR
\ENDFOR

\IF{$\varPi_{n,0} = \emptyset \land \varPi_{n,1} = \emptyset$}
    \STATE Report ``$N_0$ and $N_1$ are equivalent models.''
\ELSIF{$\varPi_{n,0} = \emptyset \land \varPi_{n,1} \neq \emptyset$}
    \STATE Report ``$N_0 \sqsubseteq N_1$ and $N_1 \not\sqsubseteq N_0$.''
\ELSIF{$\varPi_{n,0} \neq \emptyset \land \varPi_{n,1} = \emptyset$}
    \STATE Report ``$N_1 \sqsubseteq N_0$ and $N_0 \not\sqsubseteq N_1$.''
\ELSE
    \STATE Report ``Two models may not be equivalent.''
\ENDIF

\RETURN $\langle \varPi_0, \varPi_1, E, \eta_t, \varPi_{n,0}, \varPi_{n,1} \rangle$
\end{algorithmic}

\end{algorithm}
%
We demonstrate the containment checking between two Petri nets $N_0$ and $N_1$ shown in Figures~\ref{fig:PN}(c) and (d), respectively. 

\indent{\em 1.} Step 1 of ContainmentChecker (Algorithm \ref{Algo:veriDCP}) accepts the user supplied place correspondence between the place \( p_0 \) of \( N_0 \) and the place \( p'_0 \) of \( N_1 \) to start the process. 

\indent{\em 2.} Step 2 of Algorithm \ref{Algo:veriDCP} invokes PathConstructor method(Algorithm \ref{Algo:constAllPaths} in Appendix) to generate the set of paths for both $N_0$ and $N_1$ as 
$\{\alpha_1, \alpha_2, \ldots, \alpha_8\}$ and 
$\{\beta_1, \beta_2, \ldots, \beta_{12}\}$, respectively. 
The dotted boundaries for $N_0$ and $N_1$ represent these paths. The PathConstructor method invokes Algorithm \ref{Algo:Token}( in Appendix) to generate execution cut-points and paths corresponding to these cut-points. For instance, the Algorithm \ref{Algo:Token} marked the place $p_6 $ of $N_0$  and the places $p'_7$ and $p'_9$ of $N_1$ as {\em execution cut-points}. 

\indent{\em 3. (Path Extension)}: Step 6 of Algorithm \ref{Algo:veriDCP} identifies a set of candidate paths in $N_1$ for a path $\alpha$ in $N_0$. To begin with, the place correspondence between \( p_0 \) of \( N_0 \) and \( p'_0 \) of \( N_1 \) is considered. Accordingly, the path \( \alpha_1 \) of \( N_0 \) is selected as the candidate path for \( \beta_1 \) of \( N_1 \) for equivalence checking. 
Since \( R_{\alpha_1} \equiv\)  \( L_1 \rightsquigarrow C_1 \rightsquigarrow \neg C_1 \), 
and \( R_{\beta_1} \equiv\)  \( L_1 \), 
\( R_{\beta_1}\lesssim R_{\alpha_1}\) 
This is evaluated in step 17 of Algorithm \ref{Algo:veriDCP} and {\em prepareForExtension} method (Algorithm \ref{Algo:prepareForExtension} in Appendix) is called to extend \( \beta_1 \). 

Step 7 of Algorithm \ref{Algo:prepareForExtension} computes the set of post paths, i.e., the successor path of \( \beta_1 \) to be \( \beta_2 \). The execution condition of the concatenated path $(\beta_1.\beta_2)$ is \(L_1 \rightsquigarrow C_1 \rightsquigarrow S \rightsquigarrow \neg C_1\).  Since $S$ is an uncommon variable, it is dropped and the resulting \(R_{\beta_1.\beta_2} \equiv L_1 \rightsquigarrow C_1\rightsquigarrow \neg C_1\). Now $R_{\alpha_1}\simeq R_{(\beta_1.\beta_2)}$ and $r_{\alpha_1}= r_{(\beta_1.\beta_2)}$. In the next iteration, step 9 of Algorithm \ref{Algo:veriDCP} ensures \( \alpha_1 \simeq (\beta_1.\beta_2) \), and the place correspondence between \( p_3 \) and \( p'_3 \) is established.

Similarly for the path \( \alpha_2 \), step 6 of Algorithm \ref{Algo:veriDCP} selects  the candidate path \( \beta_3 \). In \( \beta_3 \), both variables \( I \) and \( J \) are initialized to 0. Therefore, variable \( I \) corresponds to variable \( J \). As a result, their execution conditions are equivalent, and the data transformations are the same. Hence, \( \alpha_2 \simeq \beta_3 \), and \( p_5 \) of \( N_0 \) corresponds to both \( p'_6 \) and \( p'_8 \) of \( N_1 \) (step 9 of Algorithm \ref{Algo:veriDCP}).

\indent{4. (Path Merging)}: The data transformation $r_\alpha$ is empty for the path \( \alpha_3 \) in $N_0$. Therefore, step 27 of Algorithm \ref{Algo:veriDCP} asserts that it is necessary to extend \( \alpha_3 \) until the data transformation becomes non empty. The post-path of \( \alpha_3 \) is \(\alpha_4\). The data transformation of the concatenated path \( \alpha_e = (\alpha_3.\alpha_4) \) is $r_{\alpha_e}=$(\texttt{a=a*10,b=b/10,I++}). Recall that the WFF constructor component (Figure~\ref{fig:funcarch}) the normalized form \cite{ds}.

In the next iteration, step 6 selects candidate paths (from $N_1$) for this new path \( \alpha_e \), to be \( \beta_4 \) and \( \beta_7 \). For the path \( \beta_4 \), the execution condition matches; however, the data transformation does not, resulting in path merging (step 20 of Algorithm \ref{Algo:veriDCP}). We now explain how path merging works.

Step 4 of Algorithm \ref{Algo:findMergedPath} first computes the pre-path of \( \beta_4 \)  which is \( \beta_3 \), and identifies the post-path of \( \beta_3 \) as \( \beta_7 \). Recall, that \( \beta_7 \) is also a candidate for \( \alpha_3 \).
Therefore, \( \beta_4 \) merges with \( \beta_7 \), forming \( \beta_m = (\beta_4 \curlyveedownarrow \beta_7) \) at step 15 of Algorithm~\ref{Algo:findMergedPath}. This process continues extending non-equivalent paths until another valid candidate path is found.
Now, $r_{\alpha_e}=r_{\beta_m}$ and \( \alpha_e \simeq \beta_m \). Therefore, place \( p_7 \) of \( N_0 \) corresponds to places \( p'_7 \) and \( p'_9 \) of \( N_1 \) which is ensure by the steps 10-14.  

Using the same approach, the path \( \alpha_5 \) is equivalent to the merged path \( \beta_6\curlyveedownarrow\beta_9 \) and \( \alpha_6 \) is equivalent to the merged path \( \beta_5\curlyveedownarrow\beta_8 \). With this informal explanation, we now formally define path merging.

\begin{definition}[Merge path]\label{d:merge}
A path \( \alpha \) is said to be a \emph{Merge path}, obtained by merging a set 
\( Q_P = \{ \alpha_1, \ldots, \alpha_k \} \) of parallelizable paths~\cite{ppl}, 
if there exists a common transition 
\( t \) such that \( t \in {}^{\circ}({}^{\circ}\alpha_i) \) 
for all \( i \), \( 1 \leq i \leq k \).
 The path $\alpha$ is denoted as ($\alpha_1 \curlyveedownarrow \ldots \curlyveedownarrow \alpha_k$). The set of pre-places of \( \alpha \) is equal to the union of the pre-places of the paths \( \alpha_1, \ldots, \alpha_k \), i.e.,
${}^{\circ}\alpha = \bigcup_{i=1}^{k} {}^{\circ}\alpha_i$.
 \end{definition}

\indent{\em 5. (One-to-one equivalence)}:
The step 9 of Algorithm \ref{Algo:veriDCP} asserts that the path \( \alpha_7 \) of \( N_0 \) are equivalent to \( \beta_{11} \) because their data transformations and execution conditions are equivalent. 
Since \texttt{WasteBin1} is uncommon in \( \beta_{10} \), it is dropped, after which step 9 also asserts that \( \alpha_8 \) is equivalent to \( \beta_{10} \).

\indent{\em 6. (Non-equivalence)}: The path \( \beta_{12} \) has no equivalent path in \( N_0 \), and its execution condition contains the uncommon variable \( S \), causing $(R_{\alpha} \neg \simeq R_{\beta})$. Therefore, the path \( \beta_{12} \) is put in $\varPi'_1$ in step 25 of Algorithm \ref{Algo:veriDCP}. 

\indent{\em 7. (Termination)}: Finally, the method identifies that the non-equivalent path sets of $N_1$ are not empty in step 30 and accordingly declares that $N_0 \sqsubseteq N_1$ (step 32), i.e., the Petri net $N_0$ is contained in $N_1$.
%


%
\paragraph*{Correctness and Complexity:}
The containment checker (Algorithm \ref{Algo:veriDCP} is iterative over a set of finite paths. At each iteration, the number of paths to be compared is reduced. Therefore, its termination is guaranteed. 
All the functions which are called by the Algorithm \ref{Algo:veriDCP}
are also terminated. The worst-case time complexity of the containment checking algorithm is exponential because
Algorithm~\ref{Algo:veriDCP} invokes the WFF constructor~\cite{ds} at steps 9, 15, 17, 20, and 21. This constructor checks the equivalence between two expressions, a process that can be exponential in the worst case.
The theorem for the soundness of Algorithm \ref{Algo:veriDCP} is given below. 
\begin{theorem}
If the function \texttt{ContainmentChecker} (Algorithm \ref{Algo:veriDCP}) 
reaches step 31 and $(a)$ returns $\varPi_{n,0} = \emptyset$, then $N_0 \sqsubseteq N_1$.
\vspace{-0.07in}
\end{theorem}

The detailed proof is given in the Appendix.


%
  \vspace{-0.1in}
\section{Experimental Results}\label{sec:expResult}
We tested the tool on the open-source OSCAT library \footnote{http://www.oscat.de/images/OSCATBasic/oscat\_basic333\_en.pdf}.
The OSCAT library comprises 1,000 applications, organized into 55 clusters based on their complexity. From this library, we selected 80 applications, each of which belongs to at least one of the 55 clusters. These 80 applications were then categorized into four classes, classified on their Lines of Code (LOC) and complexity.
The control programs are generally contained within 200 LOC. 

\begin{table*}[h]  
\centering
\begin{tabular}{|p{1.3cm}|p{11cm}|}  
\hline  
\textbf{Type} & \textbf{Description} \\  
\hline

Basic & Average 20 lines of program with one loop and two \texttt{IF THEN ELSE}.\\  
\hline 
Simple & Avg. 40 lines of program with one level nested loop and three \texttt{IF THEN ELSE}. \\
\hline
Medium & Average 60 lines with two level nested loops and there data independent loops.\\
\hline 
Complex & Average 90 lines with two or more data independent loops and two or more level nested \texttt{IF THEN ELSE}.
\\
\hline  
\end{tabular}  
\caption{Description of various types of benchmarks from OSCAT library}  
\label{T:Benchmark}
 \vspace{-0.5in}
\end{table*}

\subsection{Benchmark Preparation}
Table~\ref{T:Benchmark} provides an overview of the applications considered for our experimental study.
 We have considered 80 applications from the OSCAT library and divided them into four classes
based on their complexity.

Next, for each of the SFC programs in Table~\ref{T:Benchmark}, we have taken their upgraded versions from the OSCAT library. These upgradations are essentially the addition of sensors and actuators in the system, as well as some code improvement rules such as loop parallelization, thread-level parallelism, and code motion across the loop, to name a few. 

\subsection{Experimental Evaluation}
We used only a single core of a 2.0GHz Intel\textsuperscript{\textregistered} Core\textsuperscript{TM}2 machine to run these benchmarks. 
For both cases, we have performed the following steps systematically.

\paragraph{\bf Tool Verification:} We feed the original and the upgraded version of each benchmark as inputs to our 
tool. We evaluate the tool's performance when the original program is contained to the upgraded program and 
when the upgrade is faulty (i.e., the upgraded version does not retain the original functionality).

\paragraph{\bf Performance:} In Table \ref{tableSM},
 we report the average number of paths for both Petri net models as well as average containment checking time. 
The $4^{th}$ column of Table~\ref{tableSM} depicts whether the containment checker performed path merging (PM) to establish the containment between two models. As explained in Sec~\ref{sec:contchk}, the containment checker may have to perform path merging in the case of either loop distribution followed or thread-level parallelism or both. Here we observe that path merging takes place in the \texttt{Medium} and \texttt{Complex} classes of OSCAT benchmarks.
The $5^{th}$ column of Table~\ref{tableSM} depicts whether the containment checker performed path extension (PE) to establish the containment between two models. As explained in Sec~\ref{sec:contchk}, the containment checker may have to perform path extension in the case of either non-uniform code transformations. Here, we observe that path extension occurs in the \texttt{Simple} and \texttt{Complex} classes of OSCAT benchmarks.

\paragraph{\bf Comparison:} 
We compare our approach with a popular state-of-the-art tool \texttt{verifAPS}\footnote{\url{https://formal.kastel.kit.edu/~weigl/verifaps/stvs/}}~\cite{alexander1}, which is often used for software upgrades. This tool takes two SFCs as input and translates them into a subset of Structured Text. During this process, it eliminates all uncommon variables from the two SFCs, similar to our approach. Then they are converted to SMV models~\cite{SMV} and these two SMV models are verified for equivalence using NuXMV \footnote{https://nuxmv.fbk.eu/}. Unlike ours, \texttt{verifAPS} employs an inductive inference-based technique~\cite{mattias}. We compare our results with those produced by \texttt{verifAPS} in the last column of Table~\ref{tableSM}.
%
Our method is approximately $\approx4X$ faster than \texttt{verifAPS} because it translates the SFC to a Petri net in a syntactic way and uses path-based containment  checking. This approach eliminates the need for invariant computation at each state, unlike \texttt{verifAPS}, which translates the SFC into two intermediate representations and computes the invariant during equivalence checking.

\subsection{Fault Injection}
In order to experimentally evaluate the performance of the tool in the presence of faulty code upgradation, we inject some errors in the SFC and observe the time it takes to detect these errors during the containment checking process. We have introduced the following types of (both control level and thread 
level) erroneous code upgradation.
\begin{enumerate}
    \item {\bf Type 1:}
 non-uniform boosting up assignment statement motions from one branch of 
step to the step preceding it, which introduces false-data dependencies; this has been injected into the 
\texttt{Medium} and \texttt{Simple} benchmarks.
\item  {\bf Type 2:} non-uniform duplicating down  assignment statement motions from the step where bifurcation takes place to one 
branch that removes data dependency in the other 
branch; this has been injected in the \texttt{Complex} and \texttt{Basic} benchmarks.

\item {\bf Type 3:} data-locality transformations which introduce false 
data-locality in the body of the loop
in \texttt{Medium} and \texttt{Complex} benchmarks. 
\end{enumerate}
\begin{table}
\centering
\footnotesize
\begin{tabular}{|p{1.8cm}|p{2cm}|p{2.2cm}|r|r|r|r|}
 \hline 
 Benchmarks  & $\#$ Paths from Original SFC   & $\#$ Paths from Upgraded SFC  & PM & PE &Time (Sec) &\texttt{verifAPS} Time (sec)  \\

 \cline{1-7}
 Basic  & 7 &  7  & NO &  NO& 1.43 &5.45  \\ \hline
  Simple  & 9 & 9  & NO & YES& 1.54 &7.2 \\ \hline
Medium & 9  & 12  & YES & NO& 3.01 & 11.23 \\ \hline
Complex  & 17 & 19   & YES & YES &6.12 & 28.3 \\ \hline
\end{tabular}
\caption{Containment checking results}
\label{tableSM}
\vspace{-0.5in}
\end{table}
\begin{table}
\footnotesize
\begin{tabular}{|p{1cm}|p{2.8cm}|r|r|r|}  
\hline  
Types   & Benchmarks & Time (Sec) & 1-BisimDegree &\texttt{verifAPS} Time (Sec)\\  
\hline  
Type 1   & Medium & 12.12  & 20 \% & 55.4 \\  
   \cline{2-5}
    
    & Simple & 1.27 & 10\% & 12.2 \\
\hline  
Type 2  & Complex & 14.11 & 22\% & --\\  
     \cline{2-5} 
   & Basic & 1.23 & 8\% & -- \\
\hline  
Type 3     & Medium & 2.43 & 14\% & 10.43\\
     \cline{2-5}
     & Complex & 5.32 & 21\% & 24.31\\
   
\hline  
\end{tabular}
\caption{Non-Equivalence checking times for faulty upgradation}  
\label{T:NonEq}  
 \vspace{-0.4in}
\end{table}  

The third column of Table \ref{T:NonEq} shows the non-equivalence detection time in seconds. For OSCAT benchmarks, we have taken the average non-equivalence checking time for the ``no'' answer.
The fourth column computes the fraction of non-equivalent paths for the two programs, expressed as $1-\text{BisimDegree}$ (Definition~\ref{d:equivmetric1}). Here, the non-equivalence arises due to the injection of faults during upgradation. 

\begin{definition}[Degree of Bisimilarity]\label{d:equivmetric1}
The Degree of Bisimilarity (BisimDegree) between two models $N_0$ and $N_1$ comprising path sets $\Pi_0$ and $\Pi_1$ respectively, is defined as \(BisimDegree=\frac{|\Pi_0\cap\Pi_1|}{|\Pi_0\cup\Pi_1|}\) where $\Pi_0\cap\Pi_1$ comprises all pairs $\alpha\in\Pi_0,\beta\in\Pi_1$, such that $\alpha\simeq\beta$, as defined in Definition~\ref{d:corP}.
\end{definition}

The last column depicts the non-equivalence detection time for {\tt verifAPS} tool.  Here we observe that the tool {\tt verifAPS} takes more time to detect equivalence than our tool because of checking the invariant in each state of the translated ST code from SFC.It is also to be noted that the {\bf Type 2} error cannot be detected by {\tt verifAPS} tool. 

\begin{figure}
    \centering
    \includegraphics[width=\textwidth]{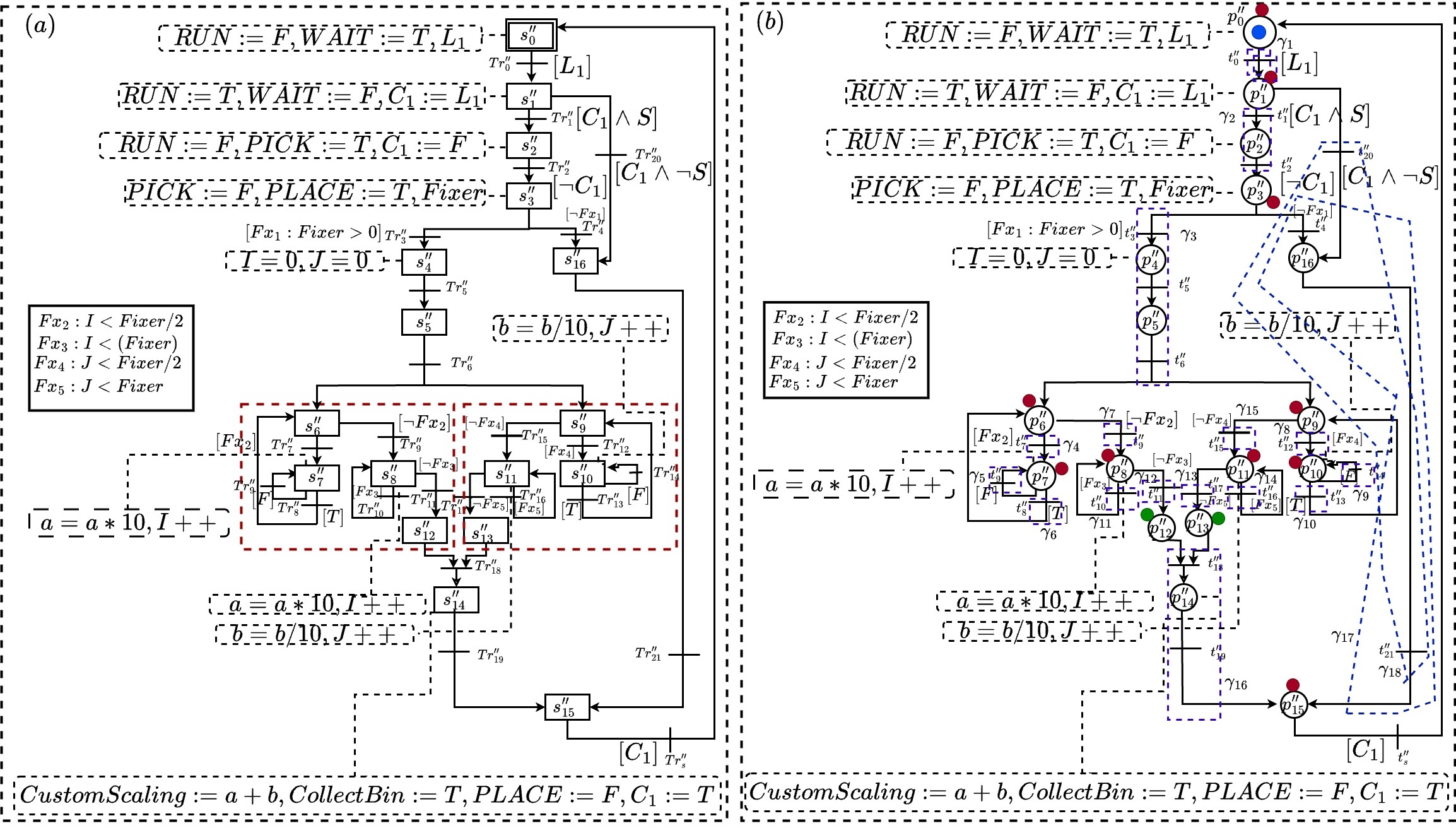}
     \caption{Limitations of our method}
    \label{fig:limitation}
   \vspace{-0.5in}
\end{figure}

\section{Discussion}\label{sec:discussion}
In light of the experimental results, we highlight several usage scenarios of the tool and mention a few limitations.

\vspace{-0.1in}
\subsection{Utility of the Tool}
\underline{\em Efficiency in Regression Testing:} Since our tool relies on symbolic containment checking, an ``yes'' answer from the tool (the original code is contained in the upgraded code) completely eliminates any further testing. However, the tool is not {\em complete}, therefore, it is always possible that the tool falsely declares two containment programs as {\em non-equivalent}. In such a case, the testing team needs to manually investigate if the upgradration process is indeed correct. Thus, the tool offers an assistive approach as it does not completely automate verification.
An objective evaluation of effort reduction requires rigorous field testing, which will be taken up by the engineering team. Currently, we have conducted a controlled field trial with a selected set of PLC experts. A ballpark comparison revealed that our verification approach is $\approx10X$ faster than manual testing; an SFC testing generally takes 3 to 5 minutes by a domain expert in a testing environment.
The experts has reported an effort reduction of the order of 60-80\% in most cases. Furthermore, we have observed that our tool has been able to successfully detect containment of nearly $90\%$ of software upgradation  use cases.

\underline{\em Upgradation Assistance:} The Developer Assistance component shown in Figure~\ref{fig:funcarch} currently generates an elaborate report of equivalent and non-equivalent paths for the old and upgraded versions with the corresponding source code. This can be easily extended, where the tool is a plugin to a popular IDE (VSCode, Eclipse) and can provide real-time feedback by highlighting the portion of the source codes that are found to be non-equivalent. The user can then decide to accept or reject the suggestion given by the tool.

\underline{\em Noteworthy Capabilities:} Our tool can handle various syntactic upgradations, including upgradation of data type, control structures, and more. Additionally, it captures several types of semantic-level upgradation, such as uniform and non-uniform code movement, loop-invariant code translation, code motion across loops, and thread-level parallelizing transformations.

\subsection{Limitations}

We illustrate an important limitation by manually creating a new version of the SFC (Figure~\ref{fig:limitation}(a)) that is functionally equivalent to the original SFC code as shown in Figure \ref{fig:SFC}(a), but non-bisimilar. We took the help of the open source SFC expert to ensure that this new version of the SFC, the original SFC (Figure \ref{fig:SFC}(a)), and the upgraded SFC (Figure~\ref{fig:SFC}(b)) are all functionally equivalent. Figure \ref{fig:limitation}(b) is the corresponding Petri net model generated by the proposed tool. Our method cannot establish containment between the Petri net in Figure~\ref{fig:limitation}(b)) and the Petri net model of the SFC code in Figure \ref{fig:PN}(c) because they are non-bisimilar.

The SFC code Figure \ref{fig:SFC}(a) has two parallel loop containing two independent operations. The variable \textit{Fixer} determines how many times the loop will run. In the hand-crafted SFC in Figure \ref{fig:limitation}(a), the operations (multiplying \texttt{a} and dividing \texttt{b}) are split into two separate loops for parallelism, and the loop is also split in half. The first loop runs for \texttt{Fixer/2} iterations, and then the second loop continues until the variable reaches \texttt{Fixer}. Since the loop iterations differ between SFC in Figure \ref{fig:SFC}(a) and the hand-crafted SFC in Figure \ref{fig:limitation}(a), the programs are considered non-bisimilar.
In Figure \ref{fig:limitation}(a), the red boundaries show the non-bisimilar portions. The containment checker can not find paths in Figure \ref{fig:PN}(c) that are equivalent to the paths $\gamma_4, \ldots \gamma_{16}$ in the 
Petri net shown in Figure \ref{fig:limitation}(b). In this example, the percentage of non-bisimilarity $(1-\text{BisimDegree})=72.2\%$.


Another type of non-bisimilarity can arise if arithmetic expressions are modified without changing the functionality. For instance, if the multiplication operation is replaced with addition and shift operations, our current version of the tool will fail to establish their equivalence. The method is not scalable at all. 

\section{Related Work}\label{sec:rw}

Verifying safety-critical systems has gained prominence since 2000 and has emerged as a serious concern across various industries. Numerous literature sources \cite{sayanMitra,safety-C-1,safety-C-2,chuchu,chuchu-1,BornotHL00,NIANG2020103328,KlikovitsLGB16,HansenTG20,GomesMDTLVM19,DBLP:phd/be/Gomes19,Bauer2004,santonus} have recognized this problem and reported various verification schemes. 
A detailed survey~\cite{article} reported different verification approaches for IEC 61131-3 languages. Simon et al. \cite{Simon2015AutomaticTC} introduced test case generation for PLC software using a model checker that iteratively created program traces, trying to increase coverage metrics. 

There have been two distinct approaches for verifying SFCs: model checking and theorem proving, both stemming from the SFC's ancestor standard Grafcet \cite{british2002grafcet}. 
The model-checking approach translates SFCs into transition systems utilizing various models such as timed automata, hybrid automata, or state machines. Several property verification techniques for PLC programs are reported in \cite{plcverify,7535242,op,FILKORN19991513,6885893,op2,DBLP:conf/etfa/RavakambinintsoaDZC24,Bauer2004,alexander1,Smet2007SafePO,NIANG2020103328,BornotHL00,DBLP:conf/cpsweek/UkegbuM23,DBLP:conf/ifm/BeckertCUVW17}. These approaches primarily focus on verifying safety or liveness properties of subsets of IEC 61131-3 languages or language features. 
On the other hand, the theorem-proving method attempts to certify the properties of SFC-based system specifications, as reported in several studies~\cite{DBLP:journals/corr/abs-1102-3529,DBLP:conf/sefm/BlechB11,DBLP:journals/corr/abs-1301-3047,Bornot2000}.

Unlike these approaches, our focus is on the behavioural verification of the upgradation process, where our tool checks whether the original code is behaviourally contained to the upgraded code.
The tool \texttt{verifAPS} reported in \cite{alexander1} describes an inductive inference-based approach for checking equivalence between two versions of SFCs.
Behavioural verification and its applications have been well studied; several techniques for code-level verification are reported in \cite{amir,ne,mit,sven,mattias,DBLP:conf/aplas/DahiyaB17,DBLP:conf/cgo/KurheKGRB22,KunduCav,KunduTcad,souban,10.1145/3368089.3409757,10.1145/3632870,10.1145/2544173.2509509,10.1007/978-3-319-94144-8_22,10.1145/3238147.3238178,10.1145/3314221.3314596}. A comprehensive and exhaustive set of benchmarks for various behavioural verification techniques is reported in \cite{9463140}.
In addition, several model-based behavioural verification methods are discussed in \cite{10.1145/1066677.1067024,DBLP:conf/ifm/OssamiJS05,10.1145/3652620.3687820,GROBELNY2012134,10.1145/3486603.3486775,9029407}.
However, to the best of our knowledge, there has been {\em no reported work} on path based behavioral verification in the context of {\em software upgradration of PLC programs}.

\vspace{-0.1in}
\section{Conclusion}\label{sec:concl}




In this work, we proposed a formally grounded verification framework to validate whether behavioral correctness is preserved during software upgrades of PLC software. Our approach introduces a novel technique that checks for behavioral containment through symbolic path equivalence of two Petri net models. 
Experimental evaluation on a diverse suite of benchmarks from the OSCAT library demonstrates the scalability and efficacy of our tool, with significant performance gains over the state-of-the-art {\em verifAPS} system. Our method is reliable, it does not give any false positive results. Our approach is language-agnostic; it can be extended to verify equivalence between any two control programs, as long as they are mapped to Petri nets. 

An important limitation of our method is, its inability to handle hierarchical SFC as well as non-bisimilar SFCs. This is due to the restriction of traversing each loop exactly once. Furthermore, our method cannot accommodate timing behaviour (like \texttt{TON} construct), writable shared variables, and only operates on integer and boolean type variables. Overcoming these limitations will be key focus areas for our {\em future work}.

\section*{Acknowledgment}
We are sincerely grateful to the reviewer for their valuable feedback, which greatly improved the presentation of this work. We also extend our heartfelt thanks to Mattias Ulbrich from KIT, Germany, for the valuable discussions that significantly enhanced this work.


\bibliographystyle{unsrt}
\bibliography{ref}
\newpage
 \section*{Appendix} 

 \begin{algorithm}[H]
\footnotesize
\caption{PathConstructor($N$)}
 $\algorithmicrequire$ A Petri net model $N$\\
 $\algorithmicensure$ Set of paths $Q$ created from the set of cut-points in $N$

 \begin{algorithmic}[1]
  \STATE $M_h \gets P_{M_0}$; /* initialized to initial marked places in $N$*/ \\
         $Q \gets \emptyset $; \\
         $T_{sh} \langle \rangle$; // Sequence of sets of transitions \\
         tick $\gets 0$;
         
 \STATE $\cal{T}\gets$\textbf{computeAllConcurTrans} $(M_h, N)$; 
  /* returns all possible set of sets of concurrent mutually exclusive transitions to $M_h$ */ \\
  \FORALL{$T_e\in \cal{T}$}
  \STATE  $Q \gets Q \bigcup$ \textbf{getPathsForexecutionCutPt} ($M_h, T_e, T_{sh},tick, N$)
\ENDFOR

 \RETURN $Q$;
  
\end{algorithmic}
 \label{Algo:constAllPaths}
\end{algorithm}

\begin{algorithm}[H]
\footnotesize
 \caption{getPathsForexecutionCutPt($M_h, T_e, T_{sh}, tick, N$)}
 $\algorithmicrequire$ 
$M_h$ is a marking,  
$T_e$ is a set of enabled maximally parallelizable transitions and $N$ is the Petri net model.
$T_{sh}$: sequence of sets of transitions.
tick: Tick value\\
$\algorithmicensure$  Set of paths $Q$\\
 \begin{algorithmic}[1]
 \STATE $C_d\gets\emptyset$;//execution cut-point list
\IF{$T_e == \emptyset$}
    \RETURN $Q$;
  \ENDIF
  \STATE $\forall t\in T_e$, mark $t$
  \STATE $M_{new} \gets T_{e}^{\circ}$ ; /* post-places of $T_e$ acquire tokens */
  
  \STATE $T_{sh} = T_{sh}.T_e$
  
  \STATE $M_h \gets (M_h-$ $^{\circ}T_{e}) \cup M_{new}$; 
    \STATE $\forall t\in T_e, \text{TickStamp}[t] \gets \text{tick}$;
   
   \IF {($\exists$ back-edge leading to some $p\in M_h$) or ($|T_{e}^{\circ}| > |T_e|)$}
  \STATE $C_d \gets C_d \cup M_h$
  \FORALL{$p' \in M_h$}
  \STATE $\alpha \gets$ \textbf{constOnePath} ($\{p'\}, T_{sh}, N$);
     // Traverse backward from $\{p'\}$ along $T_{sh}$ to construct a path up to some cut-points \\
  \STATE $Q = Q \cup \{\alpha\}$   
      \ENDFOR
 \ENDIF

  \STATE $\cal{T}\gets$  \textbf{computeAllConcurTrans} $(M_h, N)$;\\
  
  \IF{$(\cal{T} = \emptyset)$ and ($M_h \neq \emptyset$)}
   \STATE Report as invalid PRES+ Model
  \ELSE
   \FORALL{$T_e \in \cal{T}$}
      \STATE $T_e \gets T_e - \{$marked transitions of $T_e\}$  \\
        \STATE $Q\gets Q \cup$  \textbf{getPathsForexecutionCutPt} ($M_h, T_e, T_{sh}, tick+1, N$)
    \ENDFOR
    \ENDIF
    \RETURN $Q$;
\end{algorithmic}
 \label{Algo:Token}
\end{algorithm}

\begin{algorithm}[H]
\footnotesize
\caption{prepareForExtension($\gamma, \varPi^{'}, \varPi_{n}, \eta_t, E$)}
\label{Algo:prepareForExtension}
\begin{algorithmic}[1]
\REQUIRE 1. $\gamma$: a path whose extension is sought.
2. $\varPi^{'}$: a set of paths remaining from the original path cover. 
3. $\varPi_{n}$: a set of non-equivalent paths 
4. $\eta_t$: the set of corresponding transitions pairs.
5. $E:$ pair of equivalent paths of $N_0$ and $N_1$.
\ENSURE Modified $\varPi^{'}$ remaining from the original path and modified $\varPi$

\STATE $\varPi^{'} = \varPi^{'} - \{\gamma\}$; /* $\gamma$ has to be extended */

\STATE $\Gamma_{E}^{+} =$ {\bf findPostPaths} ($\gamma, \varPi^{'}$);\\
/* Computes post-paths of $\gamma$ through which $\gamma$ can be extended. 
Such paths include those which emanate from the post-place $\gamma^{\circ}$ 
(under different guards) or those which emanate from the post-places of the last transition of $\gamma$.*/

\FOR {each $\gamma^{'} \in \Gamma_{E}^{+}$}
  \STATE $\chi_{\gamma} = ${\bf findSetOfSetsOfPrePaths} ($\gamma, \gamma^{'}, \varPi^{'},E$); \\
  /*Computes all the pre-paths of $\gamma^{'}$ 
  other than $\gamma$. Some of these may not execute
  in parallel (with $\gamma$). For example, let $\{\gamma, \gamma_1, \gamma_2, \gamma_3\}$ 
  be three pre-paths of $\gamma^{'}$; assume that $\gamma_2$
  and $\gamma_3$ have an identical post-place. Hence, $\gamma_2$ and $\gamma_3$ 
  cannot execute in parallel because the models are one(k)-safe. So the 
  pre-paths (including $\gamma$) will be $\{\{\gamma, \gamma_1, \gamma_2\},~~\{\gamma, \gamma_1, \gamma_3\}\}$.*/

  \STATE $\varPi^{'} = \varPi^{'} - \{\gamma^{'}\}$;
  
  \FORALL {$\Gamma_{P} \in \chi_{\gamma}$}
    \STATE $(\Gamma'_{P},\varPi_n) = $ {\bf trimPrePaths} 
   ($ \gamma, \Gamma_{P}, \varPi^{'}, \eta_t, E$); ~~
   /* $\Gamma_{P}$ is trimmed of the members (other than $\gamma$) 
   which are found to have equivalence (without any extension) with some path in the other model. 
   If it is detected that such a path may have to be extended
   before its equivalence is found, then no action is initiated because they are already under consideration for extension.
   However, if it is found that the path does not need any further consideration (such as extension), 
   it is put in the set $\varPi_n$ which may have no equivalent path in the other model. The set is not used for extension.  */

    \IF{($\Gamma'_{P} \neq \emptyset$)}
      \STATE $\gamma_e = $ {\bf extend} ($\gamma, \Gamma'_{P}, \gamma^{'}$); 
      /*  Constructs an extended path $\gamma_e$ of the form $\Gamma_P.\gamma^{'}$.
      The function computes those pre-places of $\gamma^{'}$ paths leading to which have been found to 
      have equivalent paths and hence do not occur in $\Gamma_P$. Then the function computes the 
      pre-places of $\gamma_e$.
      In the next two steps, the function obtains the 
      post-places of $\gamma_e$ as those of $\gamma^{'}$ and 
      the last transition of $\gamma_e$ as that of $\gamma^{'}$. After that, by method of substitution the function 
      computes the condition $R_{\gamma_e}$ of
      execution and the data transformation $r_{\gamma_e}$ along the extended path $\gamma_e$.  Removes all uncommon variables from path.
      Finally, it returns the extended path $\gamma_e$ */

      \STATE $\varPi^{'} = \left(\varPi^{'} - \Gamma_{P} \right) \cup \{\gamma_e\}$;
    \ENDIF
  \ENDFOR
\ENDFOR

\RETURN $\varPi^{'}$;
\end{algorithmic}
\end{algorithm}

\begin{algorithm}[H]
\footnotesize
\caption{prepareForMerging($  \Gamma$)} 
\label{Algo:findMergedPath}
\begin{algorithmic}[1]
\REQUIRE 

$\Gamma$: A candidate set of paths such that $\Gamma \subseteq \varPi_0$ or $\Gamma \subseteq \varPi_1$

\ENSURE 
Returns a merged path $\gamma_m$ formed from converging paths (if found), or $\emptyset$ otherwise.

\STATE $\mathcal{C} \gets \Gamma$; 
/* Candidate set for merging from one model only */

    \STATE $\mathcal{P} \gets \emptyset$;
    \FORALL{path $\gamma \in \mathcal{C}$}
        \STATE $\text{PrePaths}\gets {}^{\circ}\gamma$;
        \STATE $\mathcal{P} \leftarrow \mathcal{P} \cup \text{PrePaths}$;
    \ENDFOR

    \STATE $\mathcal{S} \leftarrow$ \textbf{allNontrivialSubsets}$(\mathcal{P})$; 
    /* All subsets of size $\geq 2$ */

    \FORALL{subset $\mathcal{G} \in \mathcal{S}$}
        \STATE $\mathcal{T}_{last} \leftarrow \{ last(\gamma) \mid \gamma \in \mathcal{G} \}$;
        \IF{ $\mathcal{T}_{last}$ is singleton }
            \STATE Let $\tau$ be the unique transition in $\mathcal{T}_{last}$;
         \STATE Let $\tau^{\circ}$ be the post-places of $\tau$;
\STATE Let $P_G$ be the set of starting places of all paths in $\mathcal{G}$;

            \IF{$\exists t \mid P_G \subseteq t^{\circ}$} 
                \STATE $\gamma_m \leftarrow$ \textbf{mergePaths}$(\mathcal{G}, \tau)$; /* This condition implies that all paths in $\mathcal{G}$ originate from the transition $t$, and their execution conditions are same. */\\
                \RETURN $\gamma_m$;
            \ENDIF
        \ENDIF
    \ENDFOR

    \IF{$\mathcal{P} = \mathcal{C}$}
        \RETURN $\emptyset$; /* No further expansion, no merge point */
    \ENDIF


\end{algorithmic}
\end{algorithm}

\newpage

 {\bf Theorem 1:} A Petri net model $N_0$ is contained in another
Petri net model $N_1$, denoted as $N_0 \sqsubseteq N_1$, for all finite path cover $\varPi_0 =
\{\alpha_{0}, \alpha_{1}, \ldots, \alpha_{l}\}$ of $N_0$ for which there exists a set $\varPi_1 =
\{\beta_{0}, \beta_{1}, \ldots, \beta_{l}\}$ of paths of $N_1$ such that for all $i, 0 \leq i \leq l,$
$(i)$ $\alpha_{i} \simeq \beta_{i}$, $(ii)$ the places in $^{\circ}\alpha_{i}$ have correspondence with those in $^{\circ}\beta_{i}$
and $(iii)$ the places in $\alpha_{i}^{\circ}$ have correspondence with those in 
 $\beta_{i}^{\circ}$.

\begin{proof}
Consider any computation $\mu_{p}$ for an out-port $p$ of $N_0$. It is required to prove that for the 
out-port $p'=f_{out}(p)$
of $N_1$, there exists a computation $\mu_{p} \simeq \mu_{p'}$.
Let $\mu_{p} = \langle T_1, T_2, \ldots, T_i, \ldots, T_l\rangle$ where, $^{\circ}T_1 = P_{M_0}$, $p \in T_{l}^{\circ}$ and 
for all $i, 1 \leq i \leq l$, if $T_{i}^{\circ} \subseteq P_{M_{i}}$, for some marking $M_i$ and 
$T_{i+1}^{\circ} \subseteq P_{M_{i+1}}$,
for some marking $M_{i+1}$, then $M_{i+1}=M_{i}^{+}$. Since $\varPi_0$ is a path cover of $N_0$, $\mu_{p}$ can be 
captured as a concatenation $(\alpha_{1}^{(1)} \parallel \alpha_{2}^{(1)}\parallel \ldots \parallel \alpha_{n_1}^{(1)}).
(\alpha_{1}^{(2)} \parallel \alpha_{2}^{(2)}\parallel \ldots \parallel
\alpha_{n_2}^{(2)})\ldots (\alpha_{1}^{(l)}) = \mu_{p}^{c}$, say,
of parallelisable paths of $\varPi_0$ such that $\mu_{p} \simeq \mu_{p}^{c}$. 
(Note that the last member in the concatenated sequence
must be a single path because the last set $T_l$ in $\mu_{p}$ is a singleton because of synchronizing transition.)

From $\mu_{p}^{c}$ let us construct a concatenated sequence $\mu_{p'}^{c} = 
(\beta_{1}^{(1)} \parallel \beta_{2}^{(1)}\parallel \ldots \parallel \beta_{n_1}^{(1)}).
(\beta_{1}^{(2)} \parallel \beta_{2}^{(2)}\parallel \ldots \parallel \beta_{n_2}^{(2)})\ldots (\beta_{1}^{(l)})$ of parallelisable
paths of $N_1$ such that for all $i, 1 \leq i \leq t$, for all $j, 1 \leq j \leq n_i, \alpha_{j}^{(i)} \simeq \beta_{j}^{(i)}$ with
any place in $^{\circ}( \alpha_{j}^{(i)})$ having a correspondence with some place in $^{\circ}(\beta_{j}^{(i)})$ and 
any place in $( \alpha_{j}^{(i)})^{\circ}$ having a correspondence with some place in $(\beta_{j}^{(i)})^{\circ}$.
From the premise of the theorem, such paths exist; also, $\mu_{p}^{c} \simeq \mu_{p'}^{c}$. 

We now prove that $\mu_{p'}^{c}$ is indeed a computation.
Consider the $i^{th}$ group $(\beta_{1}^{(i)} \parallel \beta_{2}^{(i)}\parallel \ldots \parallel \beta_{n_i}^{(i)})$ of $\mu_{p'}^{c}$.
For any $j, 1 \leq j \leq n_i$, let the $j^{th}$ path $\beta_{j}^{(i)}$ in the $i^{th}$ group be the sequence 
$\langle T_{1,j}^{(i)}, T_{2,j}^{(i)}, \ldots T_{l_{j},j}^{(i)} \rangle$ of parallelisable transitions. 
For all $k, 1 \leq k \leq \max\limits_{j=1}^{n_i}(l_j)$, we combine all the $k^{th}$ transitions of all 
the paths in the $i^{th}$ group through the union operation to form a single set of parallelisable transitions $T_{k}^{(i)} = \bigcup\limits_{j=1}^{n_i}T_{k,j}^{(i)}$. 
Obviously, the paths in the $i^{th}$ group can be of varying lengths and those having lengths less than $k$ will not contribute to
the set $T_k$.
Let the maximum length of the paths occurring in the $i^{th}$ group be $m_i$. Then the above step of combining the transition sets
of the paths group-wise results in a sequence of parallelisable transitions 
$\mu_{p'}^{c'}= \langle T_{1}^{(1)}, T_{2}^{(1)},\ldots, T_{m_1}^{(1)}, 
T_{1}^{(2)},\ldots, T_{m_2}^{(2)}, \ldots, T_{1}^{(t)}, \ldots, T_{m_t}^{(t)}\rangle$. 
We show that $\mu_{p'}^{c}$ is a computation of the out-port $p'$ of $N_1$ as per definition of computation.
What remains to be proved is that for any two consecutive transition sets $T, T^{+}$ in $\mu_{p'}^{c}$, if 
$(T^{+})^{\circ} \subseteq P'_{M_{i+1}}$ 
and $(T)^{\circ} \subseteq P'_{M_{i}}$, 
then $M'_{i+1} = {M'_i}^+$, i.e.,
$P'_{M_{i+1}}={P'_{M_i}}^{+}$. Recall that 

$P'_{M_{i}^{+}} = \{p \mid p \in$ $t^{\circ}$ and 
$t \in T'_{M_{i}}\} \cup \{p \mid p \in P'_{M_{i}}$ and $p \notin$ $^{\circ}T'_{M_{i}}\} \ldots (1)$

Note that $T'^{+} = T'_{M_{i}}$, the set of enabled transitions for the marking $M'_{i}$. We give the proof of 
$P'_{M_{i+1}} \subseteq P'_{M_{i}^{+}}$; the proof of $P'_{M_{i}^{+}} \subseteq P'_{M_{i+1}}$ follows identically.
Now, consider any $p \in P'_{M_{i+1}}$. Either $p' \in (T'^{+})^{\circ}$ or $p' \notin (T'^{+})^{\circ}$.

\begin{itemize}
 \item {\it Case 1:} $p' \in (T^{+})^{\circ} \Rightarrow p \in t$, 
 for some $t' \in T'^{+}= T'_{M_{i}} \Rightarrow p' \in P'_{M_{i}^{+}}$  because 
 $p' \in$ the first subset in the Definition \ref{d:succmarking} of $P'_{M_{i}^{+}}$.
 
 \item {\it Case 2:} $p \notin (T'^{+})^{\circ}$: In this case, $p' \in P'_{M_{i+1}} \Rightarrow p \in P'_{M_{i}}$ and 
 and $p' \notin {}^{\circ}t'$, for any $t' \in T'^{+}=T'_{M_{i}}\Rightarrow p \in P'_{M_{i}}$ and $p' \notin ^{\circ}T'_{M_{i}} 
 \Rightarrow
 p' \in P'_{M_{i}^{+}}$ because it belongs to the second subset in the  Definition \ref{d:succmarking} of $P'_{M_{i}^{+}}$.
\end{itemize}
\end{proof}

{\bf Theorem 2:}
If the function \texttt{ContainmentChecker} (Algorithm \ref{Algo:veriDCP}) 
reaches step 31 and $(a)$ returns $\varPi_{n,0}$ 

\begin{proof}
 
Let
$\varPi_{0}$ gives a path cover of $N_0$.
Hence, for any out-port $p$ of $N_0$, any computation $\mu_{p}$ can be represented as a concatenation,
$Q_{0}.Q_{1}. \ldots. Q_{l}$ say, of sets of parallel paths, such that $p \in Q_{l}^{\circ}$, 
$^{\circ}Q_{0} = P_{M_0}$, $Q_{l}$ is a singleton $\{\alpha_l\}$, say, and $Q_{i}$ 
contains only paths from $\varPi_0$, $0\leq i \leq l$. Whenever a path $\alpha$ is introduced in $\varPi_{0}$ 
(only in step 11 of Algorithm \ref{Algo:veriDCP}) an entry $\langle \alpha, \beta \rangle$ is introduced in $E$ 
(with $\alpha \simeq \beta$). Hence, 
for any path $\alpha \in Q_{i}$ for some $i$, there exists a path $\beta$ of $N_1$ such that $\langle \alpha, \beta \rangle \in E$.
Hence, we can construct a concatenation, $C'_{p'}$ say, of parallel paths of $N_1$ such that
$C'_{p'}= Q'_{0}.Q'_{1}. \ldots. Q'_{l}$, where 
$Q'_{i} = \{\beta \mid \langle \alpha, \beta \rangle \in E$ and $\alpha \in Q_{i}\}$.
We show that 
\begin{enumerate}
\item {\it Case 1:} $C'_{p'}$ is a computation of $f_{out}(p)$ in $N_1$.
\item {\it Case 2:}  $C'_{p'} \simeq_c \mu_{p}$.
\end{enumerate}

\begin{description}
 \item [Proof of Case 1:] 
$C'_{p'}$ is alternatively rewritten as a sequence of sets of places, namely, 
$\langle ^{\circ}Q'_{0}, ^{\circ}Q'_{1}$, $\ldots, ^{\circ}Q'_{i}, ^{\circ}Q'_{i+1},
\ldots, ^{\circ}Q'_l, {Q'_l}^\circ\rangle$.
It is required to prove $(a) {}^{\circ}Q'_1 \subseteq P'_{M_0}$, $(b) f_{out}(p) \in {Q'_l}^{\circ}$, 
$(c) {Q'_{i+1}}^{\circ} = {({Q'_i}^{\circ})}^+, 0 \leq i \leq l$.
\begin{description}
 \item [Proof of $(a)$:] 
A pair $\langle \alpha, \beta \rangle$ of paths is put in $E$ by 
the function \textit{ContainmentChecker} (Algorithm \ref{Algo:veriDCP}) (step 11) if they are ascertained to be equivalent; the latter will examine their equivalence only if they are found to 
satisfy the property $^{\circ}\alpha = P_{M_0} \Rightarrow ^{\circ}\beta = P'_{M_0}$ and 
$^{\circ}\beta = \phi^{1}(^{\circ}\alpha)$ by 
the function \textit{SelectPathForCheckingEquivalence}. Hence, $^{\circ}Q_{0} = P_{M_0} \Rightarrow 
{}^{\circ}Q'_{0} =P'_{M_0}$ and $^{\circ}Q'_{0} = f_{in}({}^{\circ}Q_{0})$.

\item[Proof of $(b)$:] By construction of $C'_{p'}$ from $\mu_{p}$, $Q'_{l} = \{\beta_l\}$ is a singleton and if $Q_{l}=\{\alpha_l\}$, 
then $\langle \alpha_l, \beta_l \rangle$ $ \in E$
(by construction of $C'_{p'}$). Again, by similar reasoning as in the proof of $(a)$, 
we find that the function \texttt{SelectPathForCheckingEquivalence} that 
${\alpha_l}^{\circ} \in outP_0 \Rightarrow {\beta_l}^{\circ} (= {Q'_l}^{\circ}) \in f_{out}({\alpha_l}^{\circ}) = f_{out}(p)$.

\item [Proof of $(c)$:] Given $C'_{p'} = Q'_{0}.Q'_{1}. \ldots. Q'_{l}$. We construct the corresponding 
sequence of subsets of place marking
$\varrho = \langle P_{M_0}, P_{M_1}, \ldots, P_{M_l}, P_{M_{l+1}} \rangle$ such that 
\begin{equation} \label{eq:eq7}
 P_{M_0} = {}^{\circ}C'_{p'} 
\end{equation}
\vspace*{-0.4in}

\begin{align}\label{eq:eq9}
\forall i, 0 \leq i \leq l, P_{M_{i+1}} & = \{p \mid p \in {Q'_i}^{\circ} \cap {}^{\circ}Q'_{i+1}\} \ldots (a) \\
       & \cup \{p \mid p \in {Q'_i}^{\circ} - {}^{\circ}Q'_{i+1}\} \ldots (b) \\
       &        \cup \{p \mid p \in P_{M_{i}} - {}^{\circ}Q'_{i}\} \ldots (c) 
\end{align}

\begin{equation} \label{eq:eq8}
 P_{M_l} = {Q'_l}^{\circ}  
\end{equation}
Now, $C'_{p'}$ is a computation of $N_1$ if $\varrho$ is a computation of $p'$; hence it 
is required to prove 
$(I) P_{M_0} \subseteq inP_1$, $(II) P_{M_{l+1}} = \{p'\}$, $(III) P_{M_{i+1}} = P_{M_{i}^{+}}, 0 \leq i <l$.

\begin{description}
\item [Proof of $(I)$ and $(II)$:] $(I)$ and $(II)$ are already proved in part $(a)$ and part $(b)$. 

\item [Proof of $(III)$:] If $p \in P_{M_{i+1}}$ by clause \ref{eq:eq9}(a) or \ref{eq:eq9}(b), 
then $p \in {Q'_i}^{\circ}$, where $Q'_{i} = T_{M_i}$, the set of enabled 
transitions from marking $M_i$. 
If $p \in P_{M_{i+1}}$ by clause \ref{eq:eq9}(c), then 
$p \in P_{M_i}$ and $p \notin {}^{\circ}Q'_{l}$
$\Rightarrow p \in P_{M_i}$ and $p \notin T_{M_{i}}$. 
Thus, the set $P_{M_{i+1}}$ of places satisfies the Definition \ref{d:succmarking}) of place successor
marking. Hence $P_{M_{i+1}} = P_{M_{i}^{+}}$.
\end{description}
\end{description}

\item [Proof of Case 2:]
In $C'_{p'}$, $\forall p_1 \in {Q'_i}^{\circ}$ , $0 \leq i \leq l$, there exists a 
concatenated path $\gamma_{p_1}$ of the form 
${Q'_0}^{(p_1)}.{Q'_1}^{(p_1)}. \ldots. {Q'_i}^{(p_1)}$ such that $({Q'_i}^{p_1})^{\circ} = \{p_1\}$
(by Definition of concatenated paths\cite{ppl}). The path $\gamma_{p_1}$ has a condition of execution 
$R_{\gamma_{p_1}}^{Q'_{0}}(\phi(^{\circ}\gamma_{p_1}))$, the data transformation
$r_{\gamma_{p_1}}^{Q'_{0}}(\phi(^{\circ}\gamma_{p_1}))$ and the tickStamp \texttt{TickStamp(last}($\gamma_{p_1}^{Q'_{0}}$)
Similarly,  in $\mu_{0,p}$, $\forall p_0 \in {Q_i}^{\circ}$, $0 \leq i \leq l$, we can have a path $\gamma_{p_0}$. 
So, to prove that $C'_{p'} \simeq \mu_{p}$, we have to show that
$\forall i, 0 \leq i \leq l$, $\forall p_1 \in Q'{i}^{\circ}$, $\exists p_0 \in Q_{i}^{\circ}$ such that 
$\langle p_0,p_1\rangle\in \eta_p$,
$R_{\gamma_{p_1}}^{Q'_{0}}(\phi(^{\circ}\gamma_{p_1}))$ 
$\equiv$ $R_{\gamma_{p_0}}^{Q_{0}}(\phi(^{\circ}\gamma_{p_0}))$ 
,
$r_{\gamma_{p_1}}^{Q'_{0}}(\phi(^{\circ}\gamma_{p_1}))$ 
$=$ $r_{\gamma_{p_0}}^{Q_{0}}(\phi(^{\circ}\gamma_{p_0}))$ and 
\texttt{TickStamp(last}($\gamma_{p_1}^{Q'_{0}}$) = \texttt{TickStamp(last}($\gamma_{p_0}^{Q_{0}}$)
by induction
on $i$.

\begin{itemize}
 \item [Basis ($i=0$):] $\forall p_1 \in {Q'_0}^{\circ}, \exists \beta \in Q'_0$, 
 such that $\beta^{\circ} = \{p_1\}$. So, $\gamma_{p_1} = \beta$.
By construction of $C'_{p'}$ from $\mu_{p}$,
$\exists \alpha, \langle \alpha, \beta \rangle \in E$ and $\alpha \in Q_{0}$.
Let $\{p_0\}$ be $\alpha^{\circ}$; so $\alpha = \gamma_{p_0}$.
As $\langle \alpha, \beta\rangle \in E$, $R_{\alpha}(\phi(^{\circ}\alpha)$ 
$\equiv R_{\beta}(\phi(^{\circ}\beta)$ 
$r_{\alpha}(\phi(^{\circ}\alpha) = r_{\beta}(\phi(^{\circ}\beta)$ and 
\texttt{TickStamp(last}($\alpha$)) = \texttt{TickStamp(last}($\beta$))
(ensured by Algorithm \ref{Algo:veriDCP}).
Therefore, $R_{\gamma_{p_1}}^{Q'_{1}}(\phi(^{\circ}\gamma_{p_1}))$ $\equiv$ 
$R_{\gamma_{p_0}}^{Q_{0}}(\phi(^{\circ}\gamma_{p_0}))$,
$r_{\gamma_{p_1}}^{Q'_{0}}(\phi(^{\circ}\gamma_{p_1}))$ $=$ 
$r_{\gamma_{p_0}}^{Q_{0}}(\phi(^{\circ}\gamma_{p_0}))$ and 
\texttt{TickStamp(last}($\gamma_{p_1}^{Q'_{0}}$) = \texttt{TickStamp(last}($\gamma_{p_0}^{Q_{0}}$); also, 
$\langle p_0, p_1\rangle\in \eta_p$ as ensured by Algorithm \ref{Algo:veriDCP}.

\item [Induction Hypothesis:] Let $\forall i, 0 \leq i \leq k < l$, $\forall p_1 \in {Q'_i}^{\circ}, 
\exists p_0 \in {Q_i}^{\circ}$ such that the properties
$R_{\gamma_{p_1}}^{Q'_{0}}(\phi(^{\circ}\gamma_{p_1}))$ 
$\equiv$ $R_{\gamma_{p_0}}^{Q_{0}}(\phi(^{\circ}\gamma_{p_0}))$,

$r_{\gamma_{p_1}}^{Q'_{0}}(\phi(^{\circ}\gamma_{p_1}))$ $=$ 
$r_{\gamma_{p_0}}^{Q_{0}}(\phi(^{\circ}\gamma_{p_0}))$,

\texttt{TickStamp(last}($\gamma_{p_1}^{Q'_{0}}$) = \texttt{TickStamp(last}($\gamma_{p_0}^{Q_{0}}$)

and $\langle p_0,p_1\rangle\in \eta_p$ hold.

\item [Induction Step:] 
Required to prove that $\forall p_1 \in {Q'_{k+1}}^{\circ}$, $\exists p_0 \in {Q_{k+1}}^{\circ}$ such that\\
$R_{\gamma_{p_1}}^{Q'_{0}}$ $(\phi(^{\circ}\gamma_{p_1}))$ 
$\equiv$ $R_{\gamma_{p_0}}^{Q_{0}}(\phi(^{\circ}\gamma_{p_0}))$, 
$r_{\gamma_{p_1}}^{Q'_{0}}(\phi(^{\circ}\gamma_{p_1}))$ 
$=$\\
$r_{\gamma_{p_0}}^{Q_{0}}(\phi(^{\circ}\gamma_{p_0}))$,
\texttt{TickStamp(last}($\gamma_{p_1}^{Q'_{0}}$) = \texttt{TickStamp(last}($\gamma_{p_0}^{Q_{0}}$)
and 
$\langle p_0,p_1\rangle \in \eta_p$.
Let $\gamma_{p_1} = C_{1}^{''}.\beta$, where $\beta^{\circ} = \{p_1\}$ and $C_{1}^{'}$ is a set of parallelisable paths such that
$(C_{1}^{'})^{\circ} = {}^{\circ}\beta$. Let $C_{1}^{'} = \beta_1 || \beta_2 || \ldots || \beta_{t_1}$. Now, 
$\exists \alpha, \langle \alpha, \beta \rangle \in E$ (by construction of $C'_{p'}$ from $\mu_{p}$).
Therefore, as ensured by the  Algorithm \ref{Algo:veriDCP} 
$R_{\alpha}(\phi(^{\circ}\alpha)) \equiv R_{\beta}(\phi(^{\circ}\beta))$, 
$r_{\alpha}(\phi(^{\circ}\alpha)) = r_{\beta}(\phi(^{\circ}\beta))$ and $\langle \alpha^{\circ},\beta^{\circ}\rangle$. 
Let $p_0$ be $\alpha^{\circ}$.
$\gamma_{p_0}=C_{0}^{'}.\alpha$, where $C_{0}^{'}$ is a set of parallelisable paths of the form 
$\alpha_1 || \alpha_2 || \ldots ||\alpha_{t_1}$
such that $\langle \alpha_i, \beta_i \rangle \in E, 1 \leq i \leq t_1$.
Therefore, 
$R_{\gamma_{p_1}}^{Q'_{0}}(\phi(^{\circ}\gamma_{p_1}))$ \\
$\equiv 
\bigwedge_{i=1}^{t_1} R_{\beta_i}(\phi(^{\circ}\beta_{i})) \wedge$ 
$R_{\beta}(\phi(^{\circ}\beta))\{\overline{v_1} / \phi(^{\circ}\beta)\}$\\
\hspace*{0.2in} where, $\overline{v_1} = \langle r_{\beta_1}(\phi(^{\circ}\beta_1)), r_{\beta_2}(\phi(^{\circ}\beta_2)), \ldots,
r_{\beta_{t_1}}(\phi(^{\circ}\beta_{t_1})) \rangle$)\\
$\equiv$
$\bigwedge_{i=1}^{t_1} R_{\alpha_i}(\phi(^{\circ}\alpha_{i})) \wedge 
R_{\alpha}(\phi(^{\circ}\alpha))\{\overline{v_0} / \phi(^{\circ}\alpha)\}$\\
\hspace*{0.2in} where, $\overline{v_0} = \langle r_{\alpha_1}(\phi(^{\circ}\alpha_1)), 
r_{\beta_2}(\phi(^{\circ}\alpha_2)), \ldots,
r_{\alpha_{t_1}}(\phi(^{\circ}\alpha_{t_1})) \rangle$; 
\\
$\equiv R_{C_{p_0}}^{Q_{0}}(\phi(^{\circ}C_{p_1}))$. \\
\
Since by induction hypothesis
$r_{\alpha_j}(\phi(^{\circ}\alpha_j)) = r_{\beta_j}(\phi(^{\circ}\beta_j))$,\\
\texttt{TickStamp(last}($\alpha_j$)) = \texttt{TickStamp(last}($\beta_j$))
\hspace*{0.2in}  $1 \leq j \leq t_1$, and 
$R_{\alpha_i}(\phi(^{\circ}\alpha_i))\equiv R_{\beta_i}(\phi(^{\circ}\beta_i))$,\\
since
$R_{\beta}(\phi(^{\circ}\beta)) \equiv R_{\alpha}(\phi(^{\circ}\alpha))$)\\

Similarly,
$r_{\gamma_{p_1}}^{Q'_{0}}(\phi (^{\circ}\gamma_{p_1}))$\\
$=r_{\beta}(\phi(^{\circ}\beta))\{\overline{v_1} / \phi(^{\circ}\beta)\}$\\
$= r_{\alpha}(\phi(^{\circ}\alpha))\{\overline{v_0} / \phi(^{\circ}\alpha)\}$\\
\hspace*{0.2in} since, $\langle \alpha, \beta \rangle \in E$ and $\overline{v_1} = \overline{v_0}$ by induction hypothesis\\
$= r_{C_{p_0}}^{Q_{0,0}}(\phi(^{\circ}C_{p_0}))$. 

Similarly,
\texttt{TickStamp(last}($\gamma_{p_1}$)) = \texttt{TickStamp(last}($\beta$))=\texttt{TickStamp(last}($\alpha$))
since, $\langle \alpha, \beta \rangle \in E$ and $\overline{v_1} = \overline{v_0}$ by induction hypothesis\\
= \texttt{TickStamp(last}$({C_{p_0}}^{Q_{0,0}}))$. 

\end{itemize}
\end{description}
\end{proof}

\end{document}